# New Complete Orthonormal Basis Sets of Relativistic Exponential Type Spinor Orbitals and Slater Spinor Functions of Particles with Arbitrary Half-Integral Spin in Position, Momentum and Four-Dimensional Spaces

I.I.Guseinov

*Department of Physics, Faculty of Arts and Sciences, Onsekiz Mart University, Canakkale, Turkey*

**Abstract**

Using the complete orthonormal sets of radial parts of nonrelativitistic $\psi^{\alpha}$-exponential type orbitals $(\alpha = 2,1,0,-1,-2,...)$ and spinor type tensor spherical harmonics of rank s the new formulae for the 2(2s+1)-component relativistic spinors useful in the quantum mechanical description of the arbitrary half-integral spin particles by the generalized Dirac equation introduced by the author are established in position, momentum and four-dimensional spaces, where $s = 1/2, 3/2, 5/2,...$ . These spinors are complete without the inclusion of the continuum. The 2(2s+1)-component spinors obtained are reduced to the independent sets of two-component spinors defined as a product of complete orthonormal sets of radial parts of $\psi^{\alpha}$-orbitals and two-component spinor type tensor spherical harmonics. We notice that the new idea presented in this work is the unified treatment of half-integral spin and scalar particles in position, momentum and four-dimensional spaces. Relations presented in this study can be useful in the linear combination of atomic orbitals approximation for the solution of different problems arising in the relativistic quantum mechanics when the orthonormal basis sets of relativistic exponential type spinor wave functions and Slater type spinor orbitals in position, momentum and four -dimensional spaces are employed.

1. Introduction

It is well known that the first higher spin equations have been proposed by Dirac in [1]. As was shown by Fierz and Pauli [2], these equations led to the inconsistencies in the presence of an external electromagnetic field. They resolved this difficulty for the special cases of $s = \frac{3}{2}$ and $s = 2$. Rarita and Schwinger [3] have suggested an alternative formulation of the theory on half-integral spin which avoids the complicated spinor formalism of Fierz and Pauli. They have developed theory of spin $\frac{3}{2}$ free particles which contains many of the features of the Dirac theory. The theory of spin-s free particles has been also developed in Refs [4-6]. All of these formalisms, which yield an adequate description of half-integral spin-s free particles, have many intrinsic contradictions and difficulties when an electromagnetic field interaction is introduced (see [7] and references therein). The generalized Dirac equation presented in our previous paper [8] for the particles with arbitrary half-integral spin is consistent and causal in the presence of an electromagnetic field interaction.

The elaboration of algorithms for the solution of the generalized Dirac equation in linear combination of atomic orbitals (LCAO) approach [9-11] necessitates progress in the development of theory for complete orthonormal basis sets of relativistic spinors of multiple orders. In [12] we have suggested the method for constructing in position, momentum and four-dimensional spaces the complete orthonormal basis sets for (2s+1)-component relativistic tensor wave functions and Slater tensor orbitals. Extending this approach to the case of spinors of multiple order and using the method set out in [13], we construct in this study the relevant complete orthonormal basis sets of 2(2s+1)-component relativistic $\Psi^{\alpha s}$-exponential type spinor orbitals ($\Psi^{\alpha s}$ -ETSO) for particles with arbitrary half-integral spin in position, momentum and four-dimensional spaces through the sets of two-component spinor type tensor spherical harmonics and radial parts of the complete orthonormal sets of nonrelativistic $\psi^{\alpha}$ -exponential type orbitals ($\psi^{\alpha}$ -ETO) [14] the angular parts of which are the scalar spherical harmonics. We notice that the radial parts of $\psi^{\alpha}$ -ETO correspond to the total centrally symmetric potential which contains the core attraction potential and the Lorentz potential of the field produced by the particle itself. The indices $\alpha$ arising from the use of total potential and occurring in the radial parts of $\psi^{\alpha}$ -ETO is the frictional quantum number [15]. Thus, the radial parts of the relativistic and nonrelativistic orbitals are the same and they are complete without the inclusion of the continuum. It should be noted that the nonrelativistic $\psi^{\alpha}$ -ETO are the special cases of $\Psi^{\alpha s}$ -ETSO for s=0, i.e., $\Psi^{\alpha 0} \equiv \psi^{\alpha}$ .

The relativistic spinors of multiple order obtained might be useful for solution of generalized Dirac equation of half-integral spin particles when the complete orthonormal relativistic $\Psi^{\alpha s}$-ETSO basis sets in **LCAO** approximation are employed. We notice that the definition of phases in this work for the scalar spherical harmonics $\left(Y^{*}_{lm_l} = Y_{l-m_l}\right)$ differs from the Condon-Shortley phases [16] by the sing factor $\left(Y^{*}_{lm_l} = i^{|m_l|+m_l} Y_{l-m_l}\right)$.

## 2. Relativistic spinor type tensor spherical harmonics

To construct in position, momentum and four-dimensional spaces the complete orthonormal basis sets of relativistic $\Psi^{\alpha s}$-ETSO and $\chi^{S}$-Slater type spinor orbitals ($\chi^{S}$-STSO) of 2(2s+1) order we introduce the following formulae for the independent spinor type tensor (STT) spherical harmonics of (2s+1) order (see Ref. [12]):

$$\Omega^{s}_{ljm}(\theta,\varphi)=\begin{bmatrix}\Omega^{s0}_{ljm}(\theta,\varphi)\\ \Omega^{s2}_{ljm}(\theta,\varphi)\\ \vdots\\ \Omega^{s,2s-3}_{ljm}(\theta,\varphi)\\ \Omega^{s,2s-1}_{ljm}(\theta,\varphi)\end{bmatrix} \qquad (1)$$

$$\Lambda^{s}_{\tilde{l}jm}(\theta,\varphi)=\begin{bmatrix}\Lambda^{s,2s-1}_{\tilde{l}jm}(\theta,\varphi)\\ \Lambda^{s,2s-3}_{\tilde{l}jm}(\theta,\varphi)\\ \vdots\\ \Lambda^{s2}_{\tilde{l}jm}(\theta,\varphi)\\ \Lambda^{s0}_{\tilde{l}jm}(\theta,\varphi)\end{bmatrix}. \qquad (2)$$

These STT spherical harmonics are eigenfunctions of operators $\hat{j}^2, \hat{j}_z, \hat{l}^2$ and $\hat{s}^2$. The two-component basis sets of STT spherical harmonics $\Omega^{s\lambda}_{ljm}(\theta,\varphi)$ and $\Lambda^{s,2s-\lambda}_{\tilde{l}jm}(\theta,\varphi)$ occurring in Eqs. (1) and (2) can be expressed through the scalar spherical harmonics:

$$\Omega^{s\lambda}_{ljm}(\theta,\varphi)=\begin{bmatrix}\eta_t a^{s\lambda}_{ljm}\beta_{m(\lambda)}Y_{lm(\lambda)}(\theta,\varphi)\\ -\eta_t a^{s,\lambda+1}_{ljm}\beta_{m(\lambda+1)}Y_{lm(\lambda+1)}(\theta,\varphi)\end{bmatrix} \qquad (3)$$

$$\Lambda^{s,2s-\lambda}_{\tilde{l}jm}(\theta,\varphi)=\begin{bmatrix}-ia^{s,2s-\lambda}_{\tilde{l}jm}\beta_{m(2s-\lambda)}Y_{\tilde{l}m(2s-\lambda)}(\theta,\varphi)\\ -ia^{s,2s-(\lambda+1)}_{\tilde{l}jm}\beta_{m(2s-(\lambda+1))}Y_{\tilde{l}m(2s-(\lambda+1))}(\theta,\varphi)\end{bmatrix}, \qquad (4)$$

where $\lambda=0,2,...,2s-1$, $l=j-\frac{1}{2}t$, $\tilde{l}=l+t=j+\frac{1}{2}t=2j-l$, $j\geq s$, $-j\leq m\leq j$, $t=2(j-l)=\pm1,\pm3,...,\pm2s$, $\eta_t=\frac{t}{|t|}$, $m_l=m(\lambda)=m-s+\lambda$, $\beta_{m(\lambda)}=(-1)^{[|m(\lambda)|-m(\lambda)]/2}$ and $a^{s\lambda}_{ljm}$ are the modified Clebsch-Gordan coefficients defined as

$$a^{s\lambda}_{ljm}=\left(lsm(\lambda)s-\lambda|lsjm\right). \qquad (5)$$

See Ref. [16] for the definition of Clebsch-Gordan coefficients $\left(lsm_l m-m_l|lsjm\right)$.

The STT spherical harmonics $\Omega^{s}_{ljm}(\theta,\varphi)$ and $\Lambda^{s}_{\tilde{l}jm}(\theta,\varphi)$ for fixed s satisfy the following orthonormality relations:

$$\int_0^{\pi}\int_0^{2\pi}\Omega^{s^\dagger}_{ljm}(\theta,\varphi)\Omega^{s}_{l'j'm'}(\theta,\varphi)\,Sin\theta d\theta d\varphi=\delta_{ll'}\delta_{jj'}\delta_{mm'} \qquad (6)$$

$$\int_0^{\pi}\int_0^{2\pi} \Lambda_{\tilde{l}jm}^{s^\dagger}(\theta,\varphi)\Lambda_{\tilde{l}'j'm'}^{s}(\theta,\varphi)Sin\theta d\theta d\varphi=\delta_{\tilde{l}\tilde{l}'}\delta_{jj'}\delta_{mm'}. \tag{7}$$

**3. Complete basis sets of relativistic $\Psi^{\alpha s}$-ETSO and $\chi^{S}$-STSO functions**

In order to construct the complete basis sets for 2(2s+1)-component relativistic spinor wave functions and Slater spinor orbitals from STT spherical harmonics and radial parts of nonrelativistic orbitals in position, momentum and four-dimensional spaces we use the method set out in a previous paper [8] and Eqs. (1)-(4). Then, we obtain for the complete basis sets in position space the following relations:

for relativistic spinor wave functions

$$\Psi_{nljm}^{\alpha s}(r,\theta,\varphi)=N_{n\tilde{l}}\begin{bmatrix}R_{nl}^{\alpha}(r)\Omega_{ljm}^{s}(\theta,\varphi)\\ R_{n\tilde{l}}^{\alpha}(r)\Lambda_{\tilde{l}jm}^{s}(\theta,\varphi)\end{bmatrix} \tag{8}$$

$$\bar{\Psi}_{nljm}^{\alpha s}(r,\theta,\varphi)=N_{n\tilde{l}}\begin{bmatrix}\bar{R}_{nl}^{\alpha}(r)\Omega_{ljm}^{s}(\theta,\varphi)\\ \bar{R}_{n\tilde{l}}^{\alpha}(r)\Lambda_{\tilde{l}jm}^{s}(\theta,\varphi)\end{bmatrix}, \tag{9}$$

for relativistic Slater spinor orbitals

$$\chi_{nljm}^{S}(r,\theta,\varphi)=N_{n\tilde{l}}\begin{bmatrix}R_{nl}(r)\Omega_{ljm}^{s}(\theta,\varphi)\\ R_{n\tilde{l}}(r)\Lambda_{\tilde{l}jm}^{s}(\theta,\varphi)\end{bmatrix}, \tag{10}$$

where $R_{n\tilde{l}}(r)=R_{n}(r)$ and

$$N_{n\tilde{l}}=\begin{cases}1/\sqrt{2} & for\ \ 0\le\tilde{l}\le n-1\\ 1 & for\ \ \tilde{l}\ge n\end{cases}. \tag{11}$$

The relativistic spinor wave functions ($K_{nljm}^{\alpha s},\bar{K}_{nljm}^{\alpha s}$) and Slater spinor orbitals $K_{nljm}^{s}$ in position, momentum and four-dimensional spaces are defined as

$$K_{nljm}^{\alpha s}\equiv\Psi_{nljm}^{\alpha s}(\zeta,\vec{r}),\Phi_{nljm}^{\alpha s}(\zeta,\vec{k}),Z_{nljm}^{\alpha s}(\zeta,\beta\theta\varphi) \tag{12}$$

$$\bar{K}_{nljm}^{\alpha s}\equiv\bar{\Psi}_{nljm}^{\alpha s}(\zeta,\vec{r}),\bar{\Phi}_{nljm}^{\alpha s}(\zeta,\vec{k}),\bar{Z}_{nljm}^{\alpha s}(\zeta,\beta\theta\varphi) \tag{13}$$

$$K_{nljm}^{s}\equiv\chi_{nljm}^{S}(\zeta,\vec{r}),U_{nljm}^{s}(\zeta,\vec{k}),V_{nljm}^{s}(\zeta,\beta\theta\varphi). \tag{14}$$

The nonrelativistic complete basis sets of orbitals $k_{nlm(\lambda)}^{\alpha}$, $\bar{k}_{nlm(\lambda)}^{\alpha}$ and $k_{nlm(\lambda)}$,the radial parts of which in position space occurr in Eqs. (8), (9) and (10), are determined through the corresponding nonrelativistic functions in position, momentum and four-dimensional spaces by

$$k_{nlm(\lambda)}^{\alpha}\equiv\psi_{nlm(\lambda)}^{\alpha}(\zeta,\vec{r}),\phi_{nlm(\lambda)}^{\alpha}(\zeta,\vec{k}),z_{nlm(\lambda)}^{\alpha}(\zeta,\beta\theta\varphi) \tag{15}$$

$$\bar{k}^{\alpha}_{nlm(\lambda)} \equiv \bar{\psi}^{\alpha}_{nlm(\lambda)}(\zeta,\vec{r}), \bar{\phi}^{\alpha}_{nlm(\lambda)}(\zeta,\vec{k}), \bar{z}^{\alpha}_{nlm(\lambda)}(\zeta,\beta\theta\varphi) \tag{16}$$

$$k_{nlm(\lambda)} \equiv \chi_{nlm(\lambda)}(\zeta,\vec{r}), u_{nlm(\lambda)}(\zeta,\vec{k}), v_{nlm(\lambda)}(\zeta,\beta\theta\varphi). \tag{17}$$

See Ref. [17] for the exact definition of functions occurring in Eqs. (12) - (17).

The orthonormality relations of relativistic spinor orbitals are determined as

$$\int K^{\alpha s\dagger}_{nljm}(\zeta,\vec{x}) \bar{K}^{\alpha s}_{n'l'j'm'}(\zeta,\vec{x})\, d\vec{x} = \delta_{nn'}\delta_{ll'}\delta_{jj'}\delta_{mm'} \tag{18}$$

$$\int K^{s\dagger}_{nljm}(\zeta,\vec{x}) K^{s}_{n'l'j'm'}(\zeta,\vec{x})\, d\vec{x} = \frac{(n+n')!}{\left[(2n)!(2n')!\right]^{1/2}} \delta_{ll'}\delta_{jj'}\delta_{mm'}. \tag{19}$$

As an example, the 4- and 8-component complete orthonormal basis sets of relativistic $\Psi^{\alpha s}$-ETSO through the nonrelativistic $\psi^{\alpha}$-ETO in position space for $1 \le n \le 4$ are given in Tables 1 and 2 for $s = \frac{1}{2}$ and $s = \frac{3}{2}$, respectively.

Using the relation $a^{0\lambda}_{ljm} = \delta_{jl}\delta_{mm_l}\delta_{\lambda 0}$ and formulae

$$\Omega^{0}_{ljm}(\theta,\varphi) = \beta_{m_l} Y_{lm_l}(\theta,\varphi) \tag{20a}$$

$$\Lambda^{0}_{\tilde{l}jm}(\theta,\varphi) = -i\beta_{m_l} Y_{lm_l}(\theta,\varphi) \tag{20b}$$

for the scalar particles it is easy to show that the relativistic spinor functions $K^{\alpha s}_{nljm}$, $\bar{K}^{\alpha s}_{nljm}$ and relativistic Slater spinor orbitals $K^{s}_{nljm}$ for particles with spin s=0 are reduced to the corresponding quantities for nonrelativistic complete basis sets in positon, momentum and four-dimensional spaces, i.e., $K^{\alpha s}_{nljm} \equiv k^{\alpha}_{nlm_l}, \bar{K}^{\alpha s}_{nljm} \equiv \bar{k}^{\alpha}_{nlm_l}$ and $K^{s}_{nljm} \equiv k_{nlm_l}$, where $s=0$, $j=l$, $t=0$, $m(\lambda) = m_l\delta_{\lambda 0}$ and $m = m_l$. Thus, the nonrelativistic and relativistic scalar particles can be also described by wave functions $K^{\alpha s}_{nljm}, \bar{K}^{\alpha s}_{nljm}$ and $K^{s}_{nljm}$ for s=t=0, $j=l$ and $m = m_l$

i.e., $$K^{\alpha 0}_{nljm} = \frac{1}{\sqrt{2}}\begin{bmatrix} 1 \\ -i \end{bmatrix} k^{\alpha}_{nlm_l} \tag{21a}$$

$$\bar{K}^{\alpha 0}_{nljm} = \frac{1}{\sqrt{2}}\begin{bmatrix} 1 \\ -i \end{bmatrix} \bar{k}^{\alpha}_{nlm_l} \tag{21b}$$

$$K^{0}_{nljm} = \frac{1}{\sqrt{2}}\begin{bmatrix} 1 \\ -i \end{bmatrix} k_{nlm_l}. \tag{21c}$$

## 4. Derivatives of relativistic spinor wave functions in position space

In this section, we evaluate the derivatives of $\Psi^{\alpha s}$-ETSO with respect to Cartesian coordinates that can be used in the solution of reduced Dirac equations when the LCAO approach is employed. For this purpose we use the $\Psi^{\alpha s}$-ETSO in the following form:

$$\Psi^{\alpha s}_{nljm} = N_{n\tilde{l}} \begin{bmatrix} \varphi^{s0} \\ \varphi^{s2} \\ \vdots \\ \varphi^{s,2s-3} \\ \varphi^{s,2s-1} \\ \chi^{s,2s-1} \\ \chi^{s,2s-3} \\ \vdots \\ \chi^{s2} \\ \chi^{s0} \end{bmatrix}. \tag{22}$$

Here $\varphi^{s\lambda}$ and $\chi^{s,2s-\lambda}$ are the two-component spinors defined by

$$\varphi^{s\lambda} = R^{\alpha}_{nl}\left(r\right)\Omega^{s\lambda}_{ljm}\left(\theta,\varphi\right) \tag{23}$$

$$\chi^{s,2s-\lambda} = R^{\alpha}_{n\tilde{l}}\left(r\right)\Lambda^{s,2s-\lambda}_{\tilde{l}jm}\left(\theta,\varphi\right), \tag{24}$$

where $\lambda = 0,2,\ldots,2s\text{-}1$.

Now we use the following relations:

$$\frac{\partial}{\partial z}(f\beta_m Y_{lm}) = \sum_{k=-1}^{1}{}' \left[\frac{df}{dr} + \left(\delta_{k,-1} - kl\right)\frac{f}{r}\right] b^{lm}_k \beta_m Y_{l+k,m} \tag{25}$$

$$\left(\frac{\partial}{\partial x} - i\frac{\partial}{\partial y}\right)\left(f\beta_m Y_{lm}\right) = \sum_{k=-1}^{1}{}' \left[\frac{df}{dr} + \left(\delta_{k,-1} - kl\right)\frac{f}{r}\right] d^{lm}_k \beta_{m-1} Y_{l+k,m-1} \tag{26}$$

$$\left(\frac{\partial}{\partial x} + i\frac{\partial}{\partial y}\right)\left(f\beta_m Y_{lm}\right) = \sum_{k=-1}^{1}{}' \left[\frac{df}{dr} + \left(\delta_{k,-1} - kl\right)\frac{f}{r}\right] c^{lm}_k \beta_{m+1} Y_{l+k,m+1}, \tag{27}$$

where $f$ is any function of the radial distance $r$ and

$$b^{lm}_k = \left[(l + m + \delta_{k1})(l - m + \delta_{k1}) / \left(2(l+1) + k\right)(2l + k)\right]^{1/2} \tag{28}$$

$$d^{lm}_k = -k\left[(l - km + 2\delta_{k1})(l - k\left(m-1\right)) / (2\left(l+1\right) + k)(2l + k)\right]^{1/2} \tag{29}$$

$$c^{lm}_k = k\left[(l + km + 2\delta_{k1})(l + k\left(m+1\right)) / \left(2(l+1) + k\right)(2l + k)\right]^{1/2} = -d^{l,-m}_k. \tag{30}$$

The symbol $\sum'$ in Eqs. (25), (26) and (27) indicates that the summation is to be performed in steps of two. These formulae can be obtained by the use of method set out in Ref.[18].

Using Eqs. (25), (26) and (27) we obtain for the derivatives of two-component spinors (23) and (24) the following relations:

$$c(\vec{\sigma}\hat{\vec{p}})\left[R_{nl}^{\alpha}\,\Omega_{ljm}^{s\lambda}(\theta,\varphi)\right] = c\hbar\sum_{k=-1}^{1}{}'\left[\frac{dR_{nl}^{\alpha}}{dr}+\left(\delta_{k,-1}-kl\right)\frac{R_{nl}^{\alpha}}{r}\right]{}_{k}\Lambda_{\tilde{l}jm}^{s,2s-\lambda}(\theta,\varphi) \tag{31}$$

$$c(\vec{\sigma}\hat{\vec{p}})\left[R_{n\tilde{l}}^{\alpha}\,\Lambda_{\tilde{l}jm}^{s,2s-\lambda}(\theta,\varphi)\right] = -c\hbar\sum_{k=-1}^{1}{}'\left[\frac{dR_{n\tilde{l}}^{\alpha}}{dr}+\left(\delta_{k,-1}-k\tilde{l}\right)\frac{R_{n\tilde{l}}^{\alpha}}{r}\right]{}_{k}\Omega_{ljm}^{s\lambda}(\theta,\varphi), \tag{32}$$

where

$${}_{k}\Omega_{ljm}^{s\lambda}(\theta,\varphi)=\begin{bmatrix}\eta_{t}\,{}_{k}A_{ljm}^{s\lambda}\beta_{m(\lambda)}Y_{l+k,m(\lambda)}(\theta,\varphi)\\ -\eta_{t}\,{}_{k}B_{ljm}^{s\lambda}\beta_{m(\lambda+1)}Y_{l+k,m(\lambda+1)}(\theta,\varphi)\end{bmatrix} \tag{33}$$

$${}_{k}\Lambda_{\tilde{l}jm}^{s,2s-\lambda}(\theta,\varphi)=\begin{bmatrix}-i\,{}_{k}C_{\tilde{l}jm}^{s\lambda}\beta_{m(\lambda+1)}Y_{\tilde{l}+k,m(\lambda+1)}(\theta,\varphi)\\ -i\,{}_{k}D_{\tilde{l}jm}^{s\lambda}\beta_{m(\lambda)}Y_{\tilde{l}+k,m(\lambda)}(\theta,\varphi)\end{bmatrix} \tag{34}$$

$${}_{k}A_{ljm}^{s\lambda} = a_{ljm}^{s\lambda}b_{k}^{lm(\lambda)} - a_{ljm}^{s,\lambda+1}d_{k}^{lm(\lambda+1)} \tag{35}$$

$${}_{k}B_{ljm}^{s\lambda} = a_{ljm}^{s,\lambda+1}b_{k}^{lm(\lambda+1)} + a_{ljm}^{s\lambda}c_{k}^{lm(\lambda)} \tag{36}$$

$${}_{k}C_{\tilde{l}jm}^{s\lambda} = a_{\tilde{l}jm}^{s,2s-(\lambda+1)}d_{k}^{\tilde{l}m(\lambda+1)} + a_{\tilde{l}jm}^{s,2s-\lambda}b_{k}^{\tilde{l}m(\lambda)} \tag{37}$$

$${}_{k}D_{\tilde{l}jm}^{s\lambda} = a_{\tilde{l}jm}^{s,2s-\lambda}c_{k}^{\tilde{l}m(\lambda)} - a_{\tilde{l}jm}^{s,2s-(\lambda+1)}b_{k}^{\tilde{l}m(\lambda+1)} \tag{38}$$

Some values of coefficients ${}_{k}A_{ljm}^{s\lambda}, {}_{k}B_{ljm}^{s\lambda}, {}_{k}C_{\tilde{l}jm}^{s\lambda}$ and ${}_{k}D_{\tilde{l}jm}^{s\lambda}$ are presented in Tables 3 and 4 for $s=\frac{1}{2}$ and $s=\frac{3}{2}$, respectively.

In the case of $s=\frac{1}{2}$, we use the following properties of functions ${}_{k}\Omega_{ljm}^{\frac{1}{2}0}(\theta,\varphi)$ and ${}_{k}\Lambda_{\tilde{l}jm}^{\frac{1}{2}0}(\theta,\varphi)$:

$${}_{k}\Omega_{ljm}^{\frac{1}{2}0}(\theta,\varphi)=\delta_{kt}\,{}_{t}\Omega_{ljm}^{\frac{1}{2}0}(\theta,\varphi) \tag{39}$$

$${}_{k}\Lambda_{\tilde{l}jm}^{\frac{1}{2}0}(\theta,\varphi)=\delta_{kt}\,{}_{t}\Lambda_{\tilde{l}jm}^{\frac{1}{2}0}(\theta,\varphi), \tag{40}$$

where

$$ {}_t\Omega^{\frac{1}{2}0}_{ljm}(\theta,\varphi)=\begin{bmatrix} t\,{}_tA^{\frac{1}{2}0}_{ljm}\beta_{m(0)}Y_{l+t,m(0)}(\theta,\varphi) \\ -t\,{}_tB^{\frac{1}{2}0}_{ljm}\beta_{m(1)}Y_{l+t,m(1)}(\theta,\varphi) \end{bmatrix} \tag{41} $$

$$ {}_t\Lambda^{\frac{1}{2}0}_{\tilde{l}jm}(\theta,\varphi)=\begin{bmatrix} -i\,{}_tC^{\frac{1}{2}0}_{\tilde{l}jm}\beta_{m(1)}Y_{\tilde{l}+t,m(1)}(\theta,\varphi) \\ -i\,{}_tD^{\frac{1}{2}0}_{\tilde{l}jm}\beta_{m(0)}Y_{\tilde{l}+t,m(0)}(\theta,\varphi) \end{bmatrix}. \tag{42} $$

Then, Eqs. (31) and (32) for $s=\frac{1}{2}$ become

$$ \begin{aligned} &c(\vec{\sigma}\hat{\vec{p}})\left[R^{\alpha}_{nl}\,\Omega^{\frac{1}{2}0}_{ljm}(\theta,\varphi)\right] \\ &= c\hbar\left[\frac{dR^{\alpha}_{nl}}{dr}+(1-\kappa)\frac{R^{\alpha}_{nl}}{r}\right]\Lambda^{\frac{1}{2}0}_{\tilde{l}jm}(\theta,\varphi) \end{aligned} \tag{43} $$

$$ \begin{aligned} &c(\vec{\sigma}\hat{\vec{p}})\left[R^{\alpha}_{n\tilde{l}}\,\Lambda^{\frac{1}{2}0}_{\tilde{l}jm}(\theta,\varphi)\right] \\ &= -c\hbar\left[\frac{dR^{\alpha}_{n\tilde{l}}}{dr}+(1+\kappa)\frac{R^{\alpha}_{n\tilde{l}}}{r}\right]\Omega^{\frac{1}{2}0}_{ljm}(\theta,\varphi), \end{aligned} \tag{44} $$

where $\Omega^{\frac{1}{2}0}_{ljm}(\theta,\varphi)={}_t\Omega^{\frac{1}{2}0}_{ljm}(\theta,\varphi)$, $\Lambda^{\frac{1}{2}0}_{\tilde{l}jm}(\theta,\varphi)={}_t\Lambda^{\frac{1}{2}0}_{\tilde{l}jm}(\theta,\varphi)$ and

$$ \kappa=t\left(j+\frac{1}{2}\right)=\begin{cases} l+1 & for\ t=+1 \\ -l & for\ t=-1 \end{cases}. \tag{45} $$

As can be seen from the formulae presented in this work, all of the 2(2s+1)-component relativistic basis spinor wave functions and Slater basis spinor orbitals are expressed through the sets of two-component basis spinors the radial parts of which are determined from the corresponding nonrelativistic basis functions defined in position, momentum and four-dimensional spaces. Thus, the expansion and one-range addition theorems established in [17] for the nonrelativistic $k^{\alpha}_{nlm_l}$ and $k_{nlm_l}$ basis sets in position, momentum and four-dimensional spaces can be also used in the case of relativistic basis spinor functions $K^{\alpha s}_{nljm}$ and $K^{s}_{nljm}$.

## 5. Conclusions

The complete orthonormal basis sets of relativistic spinor orbitals and Slater type spinor functions for the arbitrary half-integral spin particles in position, momentum and four-dimensional spaces are constructed from the product of complete sets of radial orbitals of nonrelativistic $\psi^{\alpha}-ETO$ corresponding to the centrally symmetric core attraction and Lorentz potentials and spinor type tensor spherical harmonics. Currently, the Gaussian and Slater type orbitals (GTO and STO) are the most popular basis functions used in electronic structure calculations. However, the GTO and STO are not orthogonal with respect to the principal quantum numbers that creates some difficulties in calculations. Therefore, the necessity for using in the Density Functional and Hartree-Fock theories the complete orthonormal sets of nonrelativistic $\psi^{\alpha}-ETO$ and relativistic $\Psi^{\alpha s}-ETSO$ as basis functions arises.

It is shown that the relativistic spinors obtained have the following properties:

1. The relativistic spinors possess 2(2s+1) independent components.
2. The relativistic spinors are complete without the inclusion of the continuum.
3. The 2(2s+1)-component spinors are reduced to the sets of two-component spinors.
4. The relativistic spinors with s = 0 are reduced to the nonrelativistic complete sets of orbitals.

Thus, we have described in this study the unified treatment of relativistic, quasirelativistic and nonrelativistic complete orthonormal basis sets of exponential type orbitals for arbitrary half-integral spin and scalar particles in position, momentum and four-dimensional spaces. The method here presented for developing the theory of complete orthonormal basis sets of $\Psi^{\alpha s}$-ETSO and $\chi^{s}$-STSO with arbitrary half-integral spin in particularly evident in the application to the solution of different problems of describing massive particles with half-integral spin within the framework of relativistic quantum physics or quantum physical chemistry.

**Table1.** The exponential type spinor orbitals in position space for $s=\frac{1}{2}$, $1 \le n \le 4$, $0 \le l \le n-1$, $\frac{1}{2} \le j \le \frac{7}{2}$ *and* $-j \le m \le j$

| n | $l$ | j | m | t | $\tilde{l}$ | $\tilde{\Psi}^{\alpha 1/2}_{nljm}$ |
|---|---|---|---|---|---|---|
| 1 | 0 | 1/2 | 1/2 | 1 | 1 | $\left(\psi^{\alpha}_{100} \quad 0 \quad 0 \quad 0\right)$ |
| | | | -1/2 | 1 | 1 | $\left(0 \quad -\psi^{\alpha}_{100} \quad 0 \quad 0\right)$ |
| 2 | 0 | 1/2 | 1/2 | 1 | 1 | $\left(\sqrt{1/2}\psi^{\alpha}_{200} \quad 0 \quad -i\sqrt{2/6}\,\psi^{\alpha}_{211} \quad i\sqrt{1/6}\,\psi^{\alpha}_{210}\right)$ |
| | | | -1/2 | 1 | 1 | $\left(0 \quad -\sqrt{1/2}\psi^{\alpha}_{200} \quad -i\sqrt{1/6}\,\psi^{\alpha}_{210} \quad -i\sqrt{2/6}\,\psi^{\alpha}_{21-1}\right)$ |
| | 1 | 1/2 | 1/2 | -1 | 0 | $\left(\sqrt{1/6}\,\psi^{\alpha}_{210} \quad \sqrt{1/3}\,\psi^{\alpha}_{211} \quad 0 \quad -i\sqrt{1/2}\,\psi^{\alpha}_{200}\right)$ |
| | | | -1/2 | -1 | 0 | $\left(-\sqrt{1/3}\,\psi^{\alpha}_{21-1} \quad \sqrt{1/6}\,\psi^{\alpha}_{210} \quad -i\sqrt{1/2}\,\psi^{\alpha}_{200} \quad 0\right)$ |
| | | 3/2 | 3/2 | 1 | 2 | $\left(\psi^{\alpha}_{211} \quad 0 \quad 0 \quad 0\right)$ |
| | | | 1/2 | 1 | 2 | $\left(\sqrt{2/3}\,\psi^{\alpha}_{210} \quad -\sqrt{1/3}\,\psi^{\alpha}_{211} \quad 0 \quad 0\right)$ |
| | | | -1/2 | 1 | 2 | $\left(-\sqrt{1/3}\,\psi^{\alpha}_{21-1} \quad -\sqrt{2/3}\,\psi^{\alpha}_{210} \quad 0 \quad 0\right)$ |
| | | | -3/2 | 1 | 2 | $\left(0 \quad \psi^{\alpha}_{21-1} \quad 0 \quad 0\right)$ |
| 3 | 0 | 1/2 | 1/2 | 1 | 1 | $\left(\sqrt{1/2}\,\psi^{\alpha}_{300} \quad 0 \quad -i\sqrt{1/3}\,\psi^{\alpha}_{311} \quad i\sqrt{1/6}\,\psi^{\alpha}_{310}\right)$ |
| | | | -1/2 | 1 | 1 | $\left(0 \quad -\sqrt{1/2}\,\psi^{\alpha}_{300} \quad -i\sqrt{1/6}\,\psi^{\alpha}_{310} \quad -i\sqrt{1/3}\,\psi^{\alpha}_{31-1}\right)$ |
| | 1 | 1/2 | 1/2 | -1 | 0 | $\left(\sqrt{1/6}\,\psi^{\alpha}_{310} \quad \sqrt{1/3}\,\psi^{\alpha}_{311} \quad 0 \quad -i\sqrt{1/2}\,\psi^{\alpha}_{300}\right)$ |
| | | | -1/2 | -1 | 0 | $\left(-\sqrt{1/3}\,\psi^{\alpha}_{31-1} \quad \sqrt{1/6}\,\psi^{\alpha}_{310} \quad -i\sqrt{1/2}\,\psi^{\alpha}_{300} \quad 0\right)$ |
| | | 3/2 | 3/2 | 1 | 2 | $\left(\sqrt{1/2}\,\psi^{\alpha}_{311} \quad 0 \quad -i\sqrt{2/5}\,\psi^{\alpha}_{322} \quad i\sqrt{1/10}\,\psi^{\alpha}_{321}\right)$ |
| | | | 1/2 | 1 | 2 | $\left(\sqrt{1/3}\,\psi^{\alpha}_{310} \quad -\sqrt{1/6}\,\psi^{\alpha}_{311} \quad -i\sqrt{3/10}\,\psi^{\alpha}_{321} \quad i\sqrt{1/5}\,\psi^{\alpha}_{320}\right)$ |
| | | | -1/2 | 1 | 2 | $\left(-\sqrt{1/6}\,\psi^{\alpha}_{31-1} \quad -\sqrt{1/3}\,\psi^{\alpha}_{310} \quad -i\sqrt{1/5}\,\psi^{\alpha}_{320} \quad -i\sqrt{3/10}\,\psi^{\alpha}_{32-1}\right)$ |
| | | | -3/2 | 1 | 2 | $\left(0 \quad \sqrt{1/2}\,\psi^{\alpha}_{31-1} \quad i\sqrt{1/10}\,\psi^{\alpha}_{32-1} \quad i\sqrt{4/10}\,\psi^{\alpha}_{32-2}\right)$ |
| | 2 | 3/2 | 3/2 | -1 | 1 | $\left(\sqrt{1/10}\,\psi_{321}^{\alpha}\quad \sqrt{2/5}\,\psi_{322}^{\alpha}\quad 0\quad -i\sqrt{1/2}\,\psi_{311}^{\alpha}\right)$ |
| | | | 1/2 | -1 | 1 | $\left(\sqrt{1/5}\,\psi_{320}^{\alpha}\quad \sqrt{3/10}\,\psi_{321}^{\alpha}\quad -i\sqrt{1/6}\,\psi_{311}^{\alpha}\quad -i\sqrt{1/3}\,\psi_{310}^{\alpha}\right)$ |
| | | | -1/2 | -1 | 1 | $\left(-\sqrt{3/10}\,\psi_{32-1}^{\alpha}\quad \sqrt{1/5}\,\psi_{320}^{\alpha}\quad -i\sqrt{1/3}\,\psi_{310}^{\alpha}\quad i\sqrt{1/6}\,\psi_{31-1}^{\alpha}\right)$ |
| | | | -3/2 | -1 | 1 | $\left(\sqrt{2/5}\,\psi_{32-2}^{\alpha}\quad -\sqrt{1/10}\,\psi_{32-1}^{\alpha}\quad i\sqrt{1/2}\,\psi_{31-1}^{\alpha}\quad 0\right)$ |
| | | 5/2 | 5/2 | 1 | 3 | $\left(\psi_{322}^{\alpha}\quad 0\quad 0\quad 0\right)$ |
| | | | 3/2 | 1 | 3 | $\left(\sqrt{4/5}\,\psi_{321}^{\alpha}\quad -\sqrt{1/5}\,\psi_{322}^{\alpha}\quad 0\quad 0\right)$ |
| | | | 1/2 | 1 | 3 | $\left(\sqrt{3/5}\,\psi_{320}^{\alpha}\quad -\sqrt{2/5}\,\psi_{321}^{\alpha}\quad 0\quad 0\right)$ |
| | | | -1/2 | 1 | 3 | $\left(-\sqrt{2/5}\,\psi_{32-1}^{\alpha}\quad -\sqrt{3/5}\,\psi_{320}^{\alpha}\quad 0\quad 0\right)$ |
| | | | -3/2 | 1 | 3 | $\left(\sqrt{1/5}\,\psi_{32-2}^{\alpha}\quad \sqrt{4/5}\,\psi_{32-1}^{\alpha}\quad 0\quad 0\right)$ |
| | | | -5/2 | 1 | 3 | $\left(0\quad -\psi_{32-2}^{\alpha}\quad 0\quad 0\right)$ |
| 4 | 0 | 1/2 | 1/2 | 1 | 1 | $\left(\sqrt{1/2}\,\psi_{400}^{\alpha}\quad 0\quad -i\sqrt{1/3}\,\psi_{411}^{\alpha}\quad i\sqrt{1/6}\,\psi_{410}^{\alpha}\right)$ |
| | | | -1/2 | 1 | 1 | $\left(0\quad -\sqrt{1/2}\,\psi_{400}^{\alpha}\quad -i\sqrt{1/6}\,\psi_{410}^{\alpha}\quad -i\sqrt{1/3}\,\psi_{41-1}^{\alpha}\right)$ |
| | 1 | 1/2 | 1/2 | -1 | 0 | $\left(\sqrt{1/6}\,\psi_{410}^{\alpha}\quad \sqrt{1/3}\,\psi_{411}^{\alpha}\quad 0\quad -i\sqrt{1/2}\,\psi_{400}^{\alpha}\right)$ |
| | | | -1/2 | -1 | 0 | $\left(-\sqrt{1/3}\,\psi_{41-1}^{\alpha}\quad \sqrt{1/6}\,\psi_{410}^{\alpha}\quad -i\sqrt{1/2}\,\psi_{400}^{\alpha}\quad 0\right)$ |
| | | 3/2 | 3/2 | 1 | 2 | $\left(\sqrt{1/2}\,\psi_{411}^{\alpha}\quad 0\quad -i\sqrt{2/5}\,\psi_{422}^{\alpha}\quad i\sqrt{1/10}\,\psi_{421}^{\alpha}\right)$ |
| | | | 1/2 | 1 | 2 | $\left(\sqrt{1/3}\,\psi_{410}^{\alpha}\quad -\sqrt{1/6}\,\psi_{411}^{\alpha}\quad -i\sqrt{3/10}\,\psi_{421}^{\alpha}\quad i\sqrt{1/5}\,\psi_{420}^{\alpha}\right)$ |
| | | | -1/2 | 1 | 2 | $\left(-\sqrt{1/6}\,\psi_{41-1}^{\alpha}\quad -\sqrt{1/3}\,\psi_{410}^{\alpha}\quad -i\sqrt{1/5}\,\psi_{420}^{\alpha}\quad -i\sqrt{3/10}\,\psi_{42-1}^{\alpha}\right)$ |
| | | | -3/2 | 1 | 2 | $\left(0\quad \sqrt{1/2}\,\psi_{41-1}^{\alpha}\quad i\sqrt{1/10}\,\psi_{42-1}^{\alpha}\quad i\sqrt{2/5}\,\psi_{42-2}^{\alpha}\right)$ |
| | 2 | 3/2 | 3/2 | -1 | 1 | $\left(\sqrt{1/10}\,\psi_{421}^{\alpha}\quad \sqrt{2/5}\,\psi_{422}^{\alpha}\quad 0\quad -i\sqrt{1/2}\,\psi_{411}^{\alpha}\right)$ |
| | | | 1/2 | -1 | 1 | $\left(\sqrt{1/5}\,\psi_{420}^{\alpha}\quad \sqrt{3/10}\,\psi_{421}^{\alpha}\quad -i\sqrt{1/6}\,\psi_{411}^{\alpha}\quad i\sqrt{1/3}\,\psi_{410}^{\alpha}\right)$ |
| | | | -1/2 | -1 | 1 | $\left(-\sqrt{3/10}\,\psi_{42-1}^{\alpha}\quad \sqrt{1/5}\,\psi_{420}^{\alpha}\quad -i\sqrt{1/3}\,\psi_{410}^{\alpha}\quad i\sqrt{1/6}\,\psi_{41-1}^{\alpha}\right)$ |
| | | | -3/2 | -1 | 1 | $\left(\sqrt{2/5}\,\psi_{42-2}^{\alpha}\quad -\sqrt{1/10}\,\psi_{42-1}^{\alpha}\quad i\sqrt{1/2}\,\psi_{41-1}^{\alpha}\quad 0\right)$ |
| | | 5/2 | 5/2 | 1 | 3 | $\left(\sqrt{1/2}\,\psi_{422}^{\alpha}\quad 0\quad -i\sqrt{3/7}\,\psi_{433}^{\alpha}\quad i\sqrt{1/14}\,\psi_{432}^{\alpha}\right)$ |
| | | | 3/2 | 1 | 3 | $\left(\sqrt{2/5}\,\psi_{421}^{\alpha}\quad -\sqrt{1/10}\,\psi_{422}^{\alpha}\quad -i\sqrt{5/14}\,\psi_{432}^{\alpha}\quad i\sqrt{1/7}\,\psi_{431}^{\alpha}\right)$ |
| | | | 1/2 | 1 | 3 | $\left(\sqrt{3/10}\,\psi_{420}^{\alpha}\quad -\sqrt{1/5}\,\psi_{421}^{\alpha}\quad -i\sqrt{2/7}\,\psi_{431}^{\alpha}\quad i\sqrt{3/14}\,\psi_{430}^{\alpha}\right)$ |
| | | | -1/2 | 1 | 3 | $\left(-\sqrt{1/5}\,\psi_{42-1}^{\alpha}\quad -\sqrt{3/10}\,\psi_{420}^{\alpha}\quad -i\sqrt{3/14}\,\psi_{430}^{\alpha}\quad -i\sqrt{2/7}\,\psi_{43-1}^{\alpha}\right)$ |
| | | | -3/2 | 1 | 3 | $\left(\sqrt{1/10}\,\psi_{42-2}^{\alpha}\quad \sqrt{2/5}\,\psi_{42-1}^{\alpha}\quad i\sqrt{1/7}\,\psi_{43-1}^{\alpha}\quad i\sqrt{5/14}\,\psi_{43-2}^{\alpha}\right)$ |
| | | | -5/2 | 1 | 3 | $\left(0\quad -\sqrt{1/2}\,\psi_{42-2}^{\alpha}\quad -i\sqrt{1/14}\,\psi_{43-2}^{\alpha}\quad -i\sqrt{3/7}\,\psi_{43-3}^{\alpha}\right)$ |
| | 3 | 5/2 | 5/2 | -1 | 2 | $\left(\sqrt{1/14}\,\psi_{432}^{\alpha}\quad \sqrt{3/7}\,\psi_{433}^{\alpha}\quad 0\quad -i\sqrt{1/2}\,\psi_{422}^{\alpha}\right)$ |
| | | | 3/2 | -1 | 2 | $\left(\sqrt{1/7}\,\psi^{\alpha}_{431} \quad \sqrt{5/14}\,\psi^{\alpha}_{432} \quad -i\sqrt{1/10}\,\psi^{\alpha}_{422} \quad -i\sqrt{2/5}\,\psi^{\alpha}_{421}\right)$ |
| | | | 1/2 | -1 | 2 | $\left(\sqrt{3/14}\,\psi^{\alpha}_{430} \quad \sqrt{2/7}\,\psi^{\alpha}_{431} \quad -i\sqrt{1/5}\,\psi^{\alpha}_{421} \quad -i\sqrt{3/10}\,\psi^{\alpha}_{420}\right)$ |
| | | | -1/2 | -1 | 2 | $\left(-\sqrt{2/7}\,\psi^{\alpha}_{43-1} \quad \sqrt{3/14}\,\psi^{\alpha}_{430} \quad -i\sqrt{3/10}\,\psi^{\alpha}_{420} \quad i\sqrt{1/5}\,\psi^{\alpha}_{42-1}\right)$ |
| | | | -3/2 | -1 | 2 | $\left(\sqrt{5/14}\,\psi^{\alpha}_{43-2} \quad -\sqrt{1/7}\,\psi^{\alpha}_{43-1} \quad i\sqrt{2/5}\,\psi^{\alpha}_{42-1} \quad -i\sqrt{1/10}\,\psi^{\alpha}_{42-2}\right)$ |
| | | | -5/2 | -1 | 2 | $\left(-\sqrt{3/7}\,\psi^{\alpha}_{43-3} \quad \sqrt{1/14}\,\psi^{\alpha}_{43-2} \quad -i\sqrt{1/2}\,\psi^{\alpha}_{42-2} \quad 0\right)$ |
| | | 7/2 | 7/2 | 1 | 4 | $\left(\psi^{\alpha}_{433} \quad 0 \quad 0 \quad 0\right)$ |
| | | | 5/2 | 1 | 4 | $\left(\sqrt{6/7}\,\psi^{\alpha}_{432} \quad -\sqrt{1/7}\,\psi^{\alpha}_{433} \quad 0 \quad 0\right)$ |
| | | | 3/2 | 1 | 4 | $\left(\sqrt{5/7}\,\psi^{\alpha}_{431} \quad -\sqrt{2/7}\,\psi^{\alpha}_{432} \quad 0 \quad 0\right)$ |
| | | | 1/2 | 1 | 4 | $\left(\sqrt{4/7}\,\psi^{\alpha}_{430} \quad -\sqrt{3/7}\,\psi^{\alpha}_{431} \quad 0 \quad 0\right)$ |
| | | | -1/2 | 1 | 4 | $\left(-\sqrt{3/7}\,\psi^{\alpha}_{43-1} \quad -\sqrt{4/7}\,\psi^{\alpha}_{430} \quad 0 \quad 0\right)$ |
| | | | -3/2 | 1 | 4 | $\left(\sqrt{2/7}\,\psi^{\alpha}_{43-2} \quad \sqrt{5/7}\,\psi^{\alpha}_{43-1} \quad 0 \quad 0\right)$ |
| | | | -5/2 | 1 | 4 | $\left(-\sqrt{1/7}\,\psi^{\alpha}_{43-3} \quad -\sqrt{6/7}\,\psi^{\alpha}_{43-2} \quad 0 \quad 0\right)$ |
| | | | -7/2 | 1 | 4 | $\left(0 \quad \psi^{\alpha}_{43-3} \quad 0 \quad 0\right)$ |

**Table2.** The exponential type spinor orbitals in position space for $s=\frac{3}{2}$, $1\le n\le 4,\ 0\le l\le n-1,\ \frac{3}{2}\le j\le\frac{9}{2}\ and\ -j\le m\le j$

| n | $l$ | j | m | t | $\tilde{l}$ | $\tilde{\Psi}^{\alpha 3/2}_{nljm}$ | | | | | | | |
|---|---|---|---|---|---|---|---|---|---|---|---|---|---|
| 1 | 0 | 3/2 | 3/2 | 3 | 3 | $\psi^{\alpha}_{100}$ | 0 | 0 | 0 | 0 | 0 | 0 | 0 |
| | | | 1/2 | 3 | 3 | 0 | $-\psi^{\alpha}_{100}$ | 0 | 0 | 0 | 0 | 0 | 0 |
| | | | -1/2 | 3 | 3 | 0 | 0 | $\psi^{\alpha}_{100}$ | 0 | 0 | 0 | 0 | 0 |
| | | | -3/2 | 3 | 3 | 0 | 0 | 0 | $-\psi^{\alpha}_{100}$ | 0 | 0 | 0 | 0 |
| 2 | 0 | 3/2 | 3/2 | 3 | 3 | $\psi^{\alpha}_{200}$ | 0 | 0 | 0 | 0 | 0 | 0 | 0 |
| | | | 1/2 | 3 | 3 | 0 | $-\psi^{\alpha}_{200}$ | 0 | 0 | 0 | 0 | 0 | 0 |
| | | | -1/2 | 3 | 3 | 0 | 0 | $\psi^{\alpha}_{200}$ | 0 | 0 | 0 | 0 | 0 |
| | | | -3/2 | 3 | 3 | 0 | 0 | 0 | $-\psi^{\alpha}_{200}$ | 0 | 0 | 0 | 0 |
| | 1 | 3/2 | 3/2 | 1 | 2 | $-\sqrt{3/5}\,\psi^{\alpha}_{210}$ | $-\sqrt{2/5}\,\psi^{\alpha}_{211}$ | 0 | 0 | 0 | 0 | 0 | 0 |
| | | | 1/2 | 1 | 2 | $\sqrt{2/5}\,\psi^{\alpha}_{21-1}$ | $1/\sqrt{15}\,\psi^{\alpha}_{210}$ | $2\sqrt{2/15}\,\psi^{\alpha}_{211}$ | 0 | 0 | 0 | 0 | 0 |
| | | | -1/2 | 1 | 2 | 0 | $-2\sqrt{2/15}\,\psi^{\alpha}_{21-1}$ | $1/\sqrt{15}\,\psi^{\alpha}_{210}$ | $-\sqrt{2/5}\,\psi^{\alpha}_{211}$ | 0 | 0 | 0 | 0 |
| | | | -3/2 | 1 | 2 | 0 | 0 | $\sqrt{2/5}\,\psi^{\alpha}_{21-1}$ | $-\sqrt{3/5}\,\psi^{\alpha}_{210}$ | 0 | 0 | 0 | 0 |
| | | 5/2 | 5/2 | 3 | 4 | $\psi^{\alpha}_{211}$ | 0 | 0 | 0 | 0 | 0 | 0 | 0 |
| | | | 3/2 | 3 | 4 | $\sqrt{2/5}\,\psi^{\alpha}_{210}$ | $-\sqrt{3/5}\,\psi^{\alpha}_{211}$ | 0 | 0 | 0 | 0 | 0 | 0 |
| | | | 1/2 | 3 | 4 | $-1/\sqrt{10}\,\psi^{\alpha}_{21-1}$ | $-\sqrt{3/5}\,\psi^{\alpha}_{210}$ | $\sqrt{3/10}\,\psi^{\alpha}_{211}$ | 0 | 0 | 0 | 0 | 0 |
| | | | -1/2 | 3 | 4 | 0 | $\sqrt{3/10}\,\psi^{\alpha}_{21-1}$ | $\sqrt{3/5}\,\psi^{\alpha}_{210}$ | $-1/\sqrt{10}\,\psi^{\alpha}_{211}$ | 0 | 0 | 0 | 0 |
| | | | -3/2 | 3 | 4 | 0 | 0 | $-\sqrt{3/5}\,\psi^{\alpha}_{21-1}$ | $-\sqrt{2/5}\,\psi^{\alpha}_{210}$ | 0 | 0 | 0 | 0 |
| | | | -5/2 | 3 | 4 | 0 | 0 | 0 | $\psi^{\alpha}_{21-1}$ | 0 | 0 | 0 | 0 |
| 3 | 0 | 3/2 | 3/2 | 3 | 3 | $\psi^{\alpha}_{300}$ | 0 | 0 | 0 | 0 | 0 | 0 | 0 |
| | | | 1/2 | 3 | 3 | 0 | $-\psi^{\alpha}_{300}$ | 0 | 0 | 0 | 0 | 0 | 0 |
| | | | -1/2 | 3 | 3 | 0 | 0 | $\psi^{\alpha}_{300}$ | 0 | 0 | 0 | 0 | 0 |
| | | | -3/2 | 3 | 3 | 0 | 0 | 0 | $-\psi_{300}^{\alpha}$ | 0 | 0 | 0 | 0 |
| | 1 | 3/2 | 3/2 | 1 | 2 | $-\sqrt{3/10}\,\psi_{310}^{\alpha}$ | $-1/\sqrt{5}\,\psi_{311}^{\alpha}$ | 0 | 0 | 0 | $-i/\sqrt{5}\,\psi_{322}^{\alpha}$ | $i/\sqrt{5}\,\psi_{321}^{\alpha}$ | $-i/\sqrt{10}\,\psi_{320}^{\alpha}$ |
| | | | 1/2 | 1 | 2 | $1/\sqrt{5}\,\psi_{31-1}^{\alpha}$ | $1/\sqrt{30}\,\psi_{310}^{\alpha}$ | $2/\sqrt{15}\,\psi_{311}^{\alpha}$ | 0 | $-i/\sqrt{5}\,\psi_{322}^{\alpha}$ | 0 | $i/\sqrt{10}\,\psi_{320}^{\alpha}$ | $i/\sqrt{5}\,\psi_{32-1}^{\alpha}$ |
| | | | -1/2 | 1 | 2 | 0 | $-2/\sqrt{15}\,\psi_{31-1}^{\alpha}$ | $1/\sqrt{30}\,\psi_{310}^{\alpha}$ | $-1/\sqrt{5}\,\psi_{311}^{\alpha}$ | $-i/\sqrt{5}\,\psi_{321}^{\alpha}$ | $i/\sqrt{10}\,\psi_{320}^{\alpha}$ | 0 | $-i/\sqrt{5}\,\psi_{32-2}^{\alpha}$ |
| | | | -3/2 | 1 | 2 | 0 | 0 | $1/\sqrt{30}\,\psi_{31-1}^{\alpha}$ | $-\sqrt{3/10}\,\psi_{310}^{\alpha}$ | $-i/\sqrt{10}\,\psi_{320}^{\alpha}$ | $-i/\sqrt{5}\,\psi_{32-1}^{\alpha}$ | $-i/\sqrt{5}\,\psi_{32-2}^{\alpha}$ | 0 |
| | | 5/2 | 5/2 | 3 | 4 | $\psi_{311}^{\alpha}$ | 0 | 0 | 0 | 0 | 0 | 0 | 0 |
| | | | 3/2 | 3 | 4 | $\sqrt{2/5}\,\psi_{310}^{\alpha}$ | $-\sqrt{3/5}\,\psi_{311}^{\alpha}$ | 0 | 0 | 0 | 0 | 0 | 0 |
| | | | 1/2 | 3 | 4 | $-1/\sqrt{10}\,\psi_{31-1}^{\alpha}$ | $-\sqrt{3/5}\,\psi_{310}^{\alpha}$ | $\sqrt{3/10}\,\psi_{311}^{\alpha}$ | 0 | 0 | 0 | 0 | 0 |
| | | | -1/2 | 3 | 4 | 0 | $\sqrt{3/10}\,\psi_{31-1}^{\alpha}$ | $\sqrt{3/5}\,\psi_{310}^{\alpha}$ | $-1/\sqrt{10}\,\psi_{311}^{\alpha}$ | 0 | 0 | 0 | 0 |
| | | | -3/2 | 3 | 4 | 0 | 0 | $-\sqrt{3/5}\,\psi_{31-1}^{\alpha}$ | $-\sqrt{2/5}\,\psi_{310}^{\alpha}$ | 0 | 0 | 0 | 0 |
| | | | -5/2 | 3 | 4 | 0 | 0 | 0 | $\psi_{31-1}^{\alpha}$ | 0 | 0 | 0 | 0 |
| | 2 | 3/2 | 3/2 | -1 | 1 | $-1/\sqrt{10}\,\psi_{320}^{\alpha}$ | $-1/\sqrt{5}\,\psi_{321}^{\alpha}$ | $-1/\sqrt{5}\,\psi_{322}^{\alpha}$ | 0 | 0 | 0 | $-i/\sqrt{5}\,\psi_{311}^{\alpha}$ | $i\sqrt{3/10}\,\psi_{310}^{\alpha}$ |
| | | | 1/2 | -1 | 1 | $1/\sqrt{5}\,\psi_{32-1}^{\alpha}$ | $-1/\sqrt{10}\,\psi_{320}^{\alpha}$ | 0 | $1/\sqrt{5}\,\psi_{322}^{\alpha}$ | 0 | $-2i/\sqrt{15}\,\psi_{311}^{\alpha}$ | $i/\sqrt{30}\,\psi_{310}^{\alpha}$ | $-i/\sqrt{5}\,\psi_{31-1}^{\alpha}$ |
| | | | -1/2 | -1 | 1 | $-1/\sqrt{5}\,\psi_{32-2}^{\alpha}$ | 0 | $1/\sqrt{10}\,\psi_{320}^{\alpha}$ | $1/\sqrt{5}\,\psi_{321}^{\alpha}$ | $-i/\sqrt{5}\,\psi_{311}^{\alpha}$ | $-i/\sqrt{30}\,\psi_{310}^{\alpha}$ | $-2i/\sqrt{15}\,\psi_{31-1}^{\alpha}$ | 0 |
| | | | -3/2 | -1 | 1 | 0 | $1/\sqrt{5}\,\psi_{32-2}^{\alpha}$ | $-1/\sqrt{5}\,\psi_{32-1}^{\alpha}$ | $1/\sqrt{10}\,\psi_{320}^{\alpha}$ | $-i\sqrt{3/10}\,\psi_{310}^{\alpha}$ | $-i/\sqrt{5}\,\psi_{31-1}^{\alpha}$ | 0 | 0 |
| | | 5/2 | 5/2 | 1 | 3 | $-\sqrt{3/7}\,\psi_{321}^{\alpha}$ | $-2/\sqrt{7}\,\psi_{322}^{\alpha}$ | 0 | 0 | 0 | 0 | 0 | 0 |
| | | | 3/2 | 1 | 3 | $-3\sqrt{2/35}\,\psi_{320}^{\alpha}$ | $-1/\sqrt{35}\,\psi_{321}^{\alpha}$ | $4/\sqrt{35}\,\psi_{322}^{\alpha}$ | 0 | 0 | 0 | 0 | 0 |
| | | | 1/2 | 1 | 3 | $3\sqrt{3/70}\,\psi_{32-1}^{\alpha}$ | $\sqrt{3/35}\,\psi_{320}^{\alpha}$ | $\sqrt{5/14}\,\psi_{321}^{\alpha}$ | $-\sqrt{6/35}\,\psi_{322}^{\alpha}$ | 0 | 0 | 0 | 0 |
| | | | -1/2 | 1 | 3 | $-\sqrt{6/35}\,\psi_{32-2}^{\alpha}$ | $-\sqrt{5/14}\,\psi_{32-1}^{\alpha}$ | $\sqrt{3/35}\,\psi_{320}^{\alpha}$ | $-3\sqrt{3/70}\,\psi_{321}^{\alpha}$ | 0 | 0 | 0 | 0 |
| | | | -3/2 | 1 | 3 | 0 | $4/\sqrt{35}\,\psi_{32-2}^{\alpha}$ | $1/\sqrt{35}\,\psi_{32-1}^{\alpha}$ | $-3\sqrt{2/35}\,\psi_{320}^{\alpha}$ | 0 | 0 | 0 | 0 |
| | | | -5/2 | 1 | 3 | 0 | 0 | $-2/\sqrt{7}\,\psi_{32-2}^{\alpha}$ | $\sqrt{3/7}\,\psi_{32-1}^{\alpha}$ | 0 | 0 | 0 | 0 |
| | | 7/2 | 7/2 | 3 | 5 | $\psi_{322}^{\alpha}$ | 0 | 0 | 0 | 0 | 0 | 0 | 0 |
| | | | 5/2 | 3 | 5 | $2/\sqrt{7}\,\psi_{321}^{\alpha}$ | $-\sqrt{3/7}\,\psi_{322}^{\alpha}$ | 0 | 0 | 0 | 0 | 0 | 0 |
| | | | 3/2 | 3 | 5 | $\sqrt{2/7}\,\psi_{320}^{\alpha}$ | $-2/\sqrt{7}\,\psi_{321}^{\alpha}$ | $1/\sqrt{7}\,\psi_{322}^{\alpha}$ | 0 | 0 | 0 | 0 | 0 |
| | | | 1/2 | 3 | 5 | $-2/\sqrt{35}\,\psi_{32-1}^{\alpha}$ | $-3\sqrt{2/35}\,\psi_{320}^{\alpha}$ | $2\sqrt{3/35}\,\psi_{321}^{\alpha}$ | $-1/\sqrt{35}\,\psi_{322}^{\alpha}$ | 0 | 0 | 0 | 0 |
| | | | -1/2 | 3 | 5 | $1/\sqrt{35}\,\psi^{\alpha}_{32-2}$ | $2\sqrt{3/35}\,\psi^{\alpha}_{32-1}$ | $3\sqrt{2/35}\,\psi^{\alpha}_{320}$ | $-2\sqrt{35}\,\psi^{\alpha}_{321}$ | 0 | 0 | 0 | 0 |
| | | | -3/2 | 3 | 5 | 0 | $-1/\sqrt{7}\,\psi^{\alpha}_{32-2}$ | $-2/\sqrt{7}\,\psi^{\alpha}_{32-1}$ | $-\sqrt{2/7}\,\psi^{\alpha}_{320}$ | 0 | 0 | 0 | 0 |
| | | | -5/2 | 3 | 5 | 0 | 0 | $\sqrt{3/7}\,\psi^{\alpha}_{32-2}$ | $2/\sqrt{7}\,\psi^{\alpha}_{32-1}$ | 0 | 0 | 0 | 0 |
| | | | -7/2 | 3 | 5 | 0 | 0 | 0 | $-\psi^{\alpha}_{32-2}$ | 0 | 0 | 0 | 0 |
| 4 | 0 | 3/2 | 3/2 | 3 | 3 | $1/\sqrt{2}\,\psi^{\alpha}_{400}$ | 0 | 0 | 0 | $-i\sqrt{2/7}\,\psi^{\alpha}_{433}$ | $i/\sqrt{7}\,\psi^{\alpha}_{432}$ | $-i\sqrt{2/35}\,\psi^{\alpha}_{431}$ | $i/\sqrt{70}\,\psi^{\alpha}_{430}$ |
| | | | 1/2 | 3 | 3 | 0 | $-1/\sqrt{2}\,\psi^{\alpha}_{400}$ | 0 | 0 | $-i/\sqrt{7}\,\psi^{\alpha}_{432}$ | $i\sqrt{6/35}\,\psi^{\alpha}_{431}$ | $-3i/\sqrt{70}\,\psi^{\alpha}_{430}$ | $-i\sqrt{2/35}\,\psi^{\alpha}_{43-1}$ |
| | | | -1/2 | 3 | 3 | 0 | 0 | $1/\sqrt{2}\,\psi^{\alpha}_{400}$ | 0 | $-i\sqrt{2/35}\,\psi^{\alpha}_{431}$ | $3i/\sqrt{70}\,\psi^{\alpha}_{430}$ | $i\sqrt{6/35}\,\psi^{\alpha}_{43-1}$ | $i/\sqrt{7}\,\psi^{\alpha}_{43-2}$ |
| | | | -3/2 | 3 | 3 | 0 | 0 | 0 | $-1/\sqrt{2}\,\psi^{\alpha}_{400}$ | $-i/\sqrt{70}\,\psi^{\alpha}_{430}$ | $-i\sqrt{2/35}\,\psi^{\alpha}_{43-1}$ | $-i/\sqrt{7}\,\psi^{\alpha}_{43-2}$ | $-i\sqrt{2/7}\,\psi^{\alpha}_{43-3}$ |
| | 1 | 3/2 | 3/2 | 1 | 2 | $-\sqrt{3/10}\,\psi^{\alpha}_{410}$ | $-1/\sqrt{5}\,\psi^{\alpha}_{411}$ | 0 | 0 | 0 | $-i/\sqrt{5}\,\psi^{\alpha}_{422}$ | $i/\sqrt{5}\,\psi^{\alpha}_{421}$ | $-i/\sqrt{10}\,\psi^{\alpha}_{420}$ |
| | | | 1/2 | 1 | 2 | $1/\sqrt{5}\,\psi^{\alpha}_{41-1}$ | $1/\sqrt{30}\,\psi^{\alpha}_{410}$ | $2/\sqrt{15}\,\psi^{\alpha}_{411}$ | | $-i/\sqrt{5}\,\psi^{\alpha}_{422}$ | 0 | $i/\sqrt{10}\,\psi^{\alpha}_{420}$ | $i/\sqrt{5}\,\psi^{\alpha}_{42-1}$ |
| | | | -1/2 | 1 | 2 | 0 | $-2/\sqrt{15}\,\psi^{\alpha}_{41-1}$ | $1/\sqrt{30}\,\psi^{\alpha}_{410}$ | $-1/\sqrt{5}\,\psi^{\alpha}_{411}$ | $-i/\sqrt{5}\,\psi^{\alpha}_{421}$ | $i/\sqrt{10}\,\psi^{\alpha}_{420}$ | 0 | $-i/\sqrt{5}\,\psi^{\alpha}_{42-2}$ |
| | | | -3/2 | 1 | 2 | 0 | 0 | $1/\sqrt{5}\,\psi^{\alpha}_{41-1}$ | $-\sqrt{3/10}\,\psi^{\alpha}_{410}$ | $-i/\sqrt{10}\,\psi^{\alpha}_{420}$ | $-i/\sqrt{5}\,\psi^{\alpha}_{42-1}$ | $-i/\sqrt{5}\,\psi^{\alpha}_{42-2}$ | 0 |
| | | 5/2 | 5/2 | 3 | 4 | $\psi^{\alpha}_{411}$ | 0 | 0 | 0 | 0 | 0 | 0 | 0 |
| | | | 3/2 | 3 | 4 | $\sqrt{2/5}\,\psi^{\alpha}_{410}$ | $-\sqrt{3/5}\,\psi^{\alpha}_{411}$ | 0 | 0 | 0 | 0 | 0 | 0 |
| | | | 1/2 | 3 | 4 | $-1/\sqrt{10}\,\psi^{\alpha}_{41-1}$ | $-\sqrt{3/5}\,\psi^{\alpha}_{410}$ | $\sqrt{3/10}\,\psi^{\alpha}_{411}$ | 0 | 0 | 0 | 0 | 0 |
| | | | -1/2 | 3 | 4 | 0 | $\sqrt{3/10}\,\psi^{\alpha}_{41-1}$ | $\sqrt{3/5}\,\psi^{\alpha}_{410}$ | $-1/\sqrt{10}\,\psi^{\alpha}_{411}$ | 0 | 0 | 0 | 0 |
| | | | -3/2 | 3 | 4 | 0 | 0 | $-\sqrt{3/5}\,\psi^{\alpha}_{41-1}$ | $-\sqrt{2/5}\,\psi^{\alpha}_{410}$ | 0 | 0 | 0 | 0 |
| | | | -5/2 | 3 | 4 | 0 | 0 | 0 | $\psi^{\alpha}_{41-1}$ | 0 | 0 | 0 | 0 |
| | 2 | 3/2 | 3/2 | -1 | 1 | $-1/\sqrt{10}\,\psi^{\alpha}_{420}$ | $-1/\sqrt{5}\,\psi^{\alpha}_{421}$ | $-1/\sqrt{5}\,\psi^{\alpha}_{422}$ | 0 | 0 | 0 | $-i/\sqrt{5}\,\psi^{\alpha}_{411}$ | $i\sqrt{3/10}\,\psi^{\alpha}_{410}$ |
| | | | 1/2 | -1 | 1 | $1/\sqrt{5}\,\psi^{\alpha}_{42-1}$ | $-1/\sqrt{10}\,\psi^{\alpha}_{420}$ | 0 | $1/\sqrt{5}\,\psi^{\alpha}_{422}$ | 0 | $-2i/\sqrt{15}\,\psi^{\alpha}_{411}$ | $i/\sqrt{30}\,\psi^{\alpha}_{410}$ | $-i/\sqrt{5}\,\psi^{\alpha}_{41-1}$ |
| | | | -1/2 | -1 | 1 | $-1/\sqrt{5}\,\psi^{\alpha}_{42-2}$ | 0 | $1/\sqrt{10}\,\psi^{\alpha}_{420}$ | $1/\sqrt{5}\,\psi^{\alpha}_{421}$ | $-i/\sqrt{5}\,\psi^{\alpha}_{411}$ | $-i/\sqrt{30}\,\psi^{\alpha}_{410}$ | $-2i/\sqrt{15}\,\psi^{\alpha}_{41-1}$ | 0 |
| | | | -3/2 | -1 | 1 | 0 | $1/\sqrt{5}\,\psi^{\alpha}_{42-2}$ | $-1/\sqrt{5}\,\psi^{\alpha}_{42-1}$ | $1/\sqrt{10}\,\psi^{\alpha}_{420}$ | $-i\sqrt{3/10}\,\psi^{\alpha}_{410}$ | $-i/\sqrt{5}\,\psi^{\alpha}_{41-1}$ | 0 | 0 |
| | | 5/2 | 5/2 | 1 | 3 | $-\sqrt{3/14}\,\psi^{\alpha}_{421}$ | $-\sqrt{2/7}\,\psi^{\alpha}_{422}$ | 0 | 0 | 0 | $-i\sqrt{15/14}/2\,\psi^{\alpha}_{433}$ | $i\sqrt{5/7}/2\,\psi^{\alpha}_{432}$ | $-i\sqrt{3/14}/2\,\psi^{\alpha}_{431}$ |
| | | | 3/2 | 1 | 3 | $-3/\sqrt{35}\,\psi^{\alpha}_{420}$ | $-1/\sqrt{70}\,\psi^{\alpha}_{421}$ | $2\sqrt{2/35}\,\psi^{\alpha}_{422}$ | 0 | $-3i/2\sqrt{14}\,\psi^{\alpha}_{433}$ | $-i/2\sqrt{7}\,\psi^{\alpha}_{432}$ | $i\sqrt{7/10}/2\,\psi^{\alpha}_{431}$ | $-3i/\sqrt{70}\,\psi^{\alpha}_{430}$ |
| | | | 1/2 | 1 | 3 | $3/2\sqrt{3/35}\,\psi^{\alpha}_{42-1}$ | $\sqrt{3/70}\,\psi^{\alpha}_{420}$ | $1/2\sqrt{5/7}\,\psi^{\alpha}_{421}$ | $-\sqrt{3/35}\,\psi^{\alpha}_{422}$ | $-i\sqrt{3/14}\,\psi^{\alpha}_{432}$ | $i/2\sqrt{35}\,\psi^{\alpha}_{431}$ | $i\sqrt{3/35}\,\psi^{\alpha}_{430}$ | $3i\sqrt{3/35}/2\,\psi^{\alpha}_{43-1}$ |
| | | | -1/2 | 1 | 3 | $-\sqrt{3/35}\,\psi^{\alpha}_{42-2}$ | $-1/2\sqrt{5/7}\,\psi^{\alpha}_{42-1}$ | $\sqrt{3/70}\,\psi^{\alpha}_{420}$ | $-3/2\sqrt{3/35}\,\psi^{\alpha}_{421}$ | $-3i\sqrt{3/35}/2\,\psi^{\alpha}_{431}$ | $i\sqrt{3/35}\,\psi^{\alpha}_{430}$ | $-i/2\sqrt{35}\,\psi^{\alpha}_{43-1}$ | $-i\sqrt{3/14}\,\psi^{\alpha}_{43-2}$ |
| | | | -3/2 | 1 | 3 | 0 | $2\sqrt{2/35}\,\psi^{\alpha}_{42-2}$ | $1/\sqrt{70}\,\psi^{\alpha}_{42-1}$ | $-3/\sqrt{35}\,\psi^{\alpha}_{420}$ | $-3i/\sqrt{70}\,\psi^{\alpha}_{430}$ | $-i\sqrt{7/10}/2\,\psi^{\alpha}_{43-1}$ | $-i/2\sqrt{7}\,\psi^{\alpha}_{43-2}$ | $3i/2\sqrt{14}\,\psi^{\alpha}_{43-3}$ |
| | | | -5/2 | 1 | 3 | 0 | 0 | $-\sqrt{2/7}\,\psi^{\alpha}_{42-2}$ | $\sqrt{3/14}\,\psi^{\alpha}_{42-1}$ | $i\sqrt{3/14}/2\,\psi^{\alpha}_{43-1}$ | $i\sqrt{5/7}/2\,\psi^{\alpha}_{43-2}$ | $i\sqrt{15/14}/2\,\psi^{\alpha}_{43-3}$ | 0 |
| | | 7/2 | 7/2 | 3 | 5 | $\psi^{\alpha}_{422}$ | 0 | 0 | 0 | 0 | 0 | 0 | 0 |
| | | | 5/2 | 3 | 5 | $2/\sqrt{7}\,\psi^{\alpha}_{421}$ | $-\sqrt{3/7}\,\psi^{\alpha}_{422}$ | 0 | 0 | 0 | 0 | 0 | 0 |
| | | | 3/2 | 3 | 5 | $\sqrt{2/7}\,\psi^{\alpha}_{420}$ | $-2/\sqrt{7}\,\psi^{\alpha}_{421}$ | $1/\sqrt{7}\,\psi^{\alpha}_{422}$ | 0 | 0 | 0 | 0 | 0 |
| | | | 1/2 | 3 | 5 | $-2/\sqrt{35}\,\psi^{\alpha}_{42-1}$ | $-3\sqrt{2/35}\,\psi^{\alpha}_{420}$ | $2\sqrt{3/35}\,\psi^{\alpha}_{421}$ | $-1/\sqrt{35}\,\psi^{\alpha}_{422}$ | 0 | 0 | 0 | 0 |
| | | | -1/2 | 3 | 5 | $1/\sqrt{35}\,\psi^{\alpha}_{42-2}$ | $2\sqrt{3/35}\,\psi^{\alpha}_{42-1}$ | $3\sqrt{2/35}\,\psi^{\alpha}_{420}$ | $-2/\sqrt{35}\,\psi^{\alpha}_{421}$ | 0 | 0 | 0 | 0 |
| | | | -3/2 | 3 | 5 | 0 | $-1/\sqrt{7}\,\psi^{\alpha}_{42-2}$ | $-2/\sqrt{7}\,\psi^{\alpha}_{42-1}$ | $-\sqrt{2/7}\,\psi^{\alpha}_{420}$ | 0 | 0 | 0 | 0 |
| | | | -5/2 | 3 | 5 | 0 | 0 | $\sqrt{3/7}\,\psi^{\alpha}_{42-2}$ | $2/\sqrt{7}\,\psi^{\alpha}_{42-1}$ | 0 | 0 | 0 | 0 |
| | | | -7/2 | 3 | 5 | 0 | 0 | 0 | $-\psi^{\alpha}_{42-2}$ | 0 | 0 | 0 | 0 |
| | 3 | 3/2 | 3/2 | -3 | 0 | $1/\sqrt{70}\,\psi^{\alpha}_{430}$ | $\sqrt{2/35}\,\psi^{\alpha}_{431}$ | $1/\sqrt{7}\,\psi^{\alpha}_{432}$ | $\sqrt{2/7}\,\psi^{\alpha}_{433}$ | 0 | 0 | 0 | $-i/\sqrt{2}\,\psi^{\alpha}_{400}$ |
| | | | 1/2 | -3 | 0 | $-\sqrt{2/35}\,\psi^{\alpha}_{43-1}$ | $3/\sqrt{70}\,\psi^{\alpha}_{430}$ | $\sqrt{6/35}\,\psi^{\alpha}_{431}$ | $1/\sqrt{7}\,\psi^{\alpha}_{432}$ | 0 | 0 | $-i/\sqrt{2}\,\psi^{\alpha}_{400}$ | 0 |
| | | | -1/2 | -3 | 0 | $1/\sqrt{7}\,\psi^{\alpha}_{43-2}$ | $-\sqrt{6/35}\,\psi^{\alpha}_{43-1}$ | $3/\sqrt{70}\,\psi^{\alpha}_{430}$ | $\sqrt{2/35}\,\psi^{\alpha}_{431}$ | 0 | $-i/\sqrt{2}\,\psi^{\alpha}_{400}$ | 0 | 0 |
| | | | -3/2 | -3 | 0 | $-\sqrt{2/7}\,\psi^{\alpha}_{43-3}$ | $1/\sqrt{7}\,\psi^{\alpha}_{43-2}$ | $-\sqrt{2/35}\,\psi^{\alpha}_{43-1}$ | $1/\sqrt{70}\,\psi^{\alpha}_{430}$ | $-i/\sqrt{2}\,\psi^{\alpha}_{400}$ | 0 | 0 | 0 |
| | | 5/2 | 5/2 | -1 | 2 | $-1/2\sqrt{3/14}\,\psi^{\alpha}_{431}$ | $-1/2\sqrt{5/7}\,\psi^{\alpha}_{432}$ | $-1/2\sqrt{15/14}\,\psi^{\alpha}_{433}$ | 0 | 0 | 0 | $-i\sqrt{2/7}\,\psi^{\alpha}_{422}$ | $i\sqrt{3/14}\,\psi^{\alpha}_{421}$ |
| | | | 3/2 | -1 | 2 | $-3/\sqrt{70}\,\psi^{\alpha}_{430}$ | $-1/2\sqrt{7/10}\,\psi^{\alpha}_{431}$ | $-1/2\sqrt{7}\,\psi^{\alpha}_{432}$ | $3/2\sqrt{14}\,\psi^{\alpha}_{433}$ | 0 | $-2i\sqrt{2/35}\,\psi^{\alpha}_{422}$ | $-i/\sqrt{70}\,\psi^{\alpha}_{421}$ | $3i/\sqrt{35}\,\psi^{\alpha}_{420}$ |
| | | | 1/2 | -1 | 2 | $3/2\sqrt{3/35}\,\psi^{\alpha}_{43-1}$ | $-\sqrt{3/35}\,\psi^{\alpha}_{430}$ | $1/2\sqrt{35}\,\psi^{\alpha}_{431}$ | $\sqrt{3/14}\,\psi^{\alpha}_{432}$ | $-i\sqrt{3/35}\,\psi^{\alpha}_{422}$ | $-i\sqrt{5/7}/2\,\psi^{\alpha}_{421}$ | $i\sqrt{3/70}\,\psi^{\alpha}_{420}$ | $-3i\sqrt{3/35}/2\,\psi^{\alpha}_{42-1}$ |
| | | | -1/2 | -1 | 2 | $-\sqrt{3/14}\,\psi^{\alpha}_{43-2}$ | $1/2\sqrt{35}\,\psi^{\alpha}_{43-1}$ | $\sqrt{3/35}\,\psi^{\alpha}_{430}$ | $3/2\sqrt{3/35}\,\psi^{\alpha}_{431}$ | $-3i\sqrt{3/35}/2\,\psi^{\alpha}_{421}$ | $-i\sqrt{3/70}\,\psi^{\alpha}_{420}$ | $-i\sqrt{5/7}/2\,\psi^{\alpha}_{42-1}$ | $i\sqrt{3/35}\,\psi^{\alpha}_{42-2}$ |
| | | | -3/2 | -1 | 2 | $3/2\sqrt{14}\,\psi^{\alpha}_{43-3}$ | $1/2\sqrt{7}\,\psi^{\alpha}_{43-2}$ | $-1/2\sqrt{7/10}\,\psi^{\alpha}_{43-1}$ | $3/\sqrt{70}\,\psi^{\alpha}_{430}$ | $-3i/\sqrt{35}\,\psi^{\alpha}_{420}$ | $-i/\sqrt{70}\,\psi^{\alpha}_{42-1}$ | $2i\sqrt{2/35}\,\psi^{\alpha}_{42-2}$ | 0 |
| | | | -5/2 | -1 | 2 | 0 | $-1/2\sqrt{15/14}\,\psi^{\alpha}_{43-3}$ | $1/2\sqrt{5/7}\,\psi^{\alpha}_{43-2}$ | $-1/2\sqrt{3/14}\,\psi^{\alpha}_{43-1}$ | $i\sqrt{3/14}\,\psi^{\alpha}_{42-1}$ | $i\sqrt{2/7}\,\psi^{\alpha}_{42-2}$ | 0 | 0 |
| | | 7/2 | 7/2 | 1 | 4 | $-1/\sqrt{3}\,\psi^{\alpha}_{432}$ | $-\sqrt{2/3}\,\psi^{\alpha}_{433}$ | 0 | 0 | 0 | 0 | 0 | 0 |
| | | | 5/2 | 1 | 4 | $-\sqrt{10/21}\,\psi^{\alpha}_{431}$ | $-1/\sqrt{7}\,\psi^{\alpha}_{432}$ | $2\sqrt{2/21}\,\psi^{\alpha}_{433}$ | 0 | 0 | 0 | 0 | 0 |
| | | | 3/2 | 1 | 4 | $-\sqrt{10/21}\,\psi^{\alpha}_{430}$ | 0 | $\sqrt{3/7}\,\psi^{\alpha}_{432}$ | $-\sqrt{2/21}\,\psi^{\alpha}_{433}$ | 0 | 0 | 0 | 0 |
| | | | 1/2 | 1 | 4 | $2\sqrt{2/21}\,\psi^{\alpha}_{43-1}$ | $\sqrt{2/21}\,\psi^{\alpha}_{430}$ | $\sqrt{2/7}\,\psi^{\alpha}_{431}$ | $-\sqrt{5/21}\,\psi^{\alpha}_{432}$ | 0 | 0 | 0 | 0 |
| | | | -1/2 | 1 | 4 | $-\sqrt{5/21}\,\psi^{\alpha}_{43-2}$ | $-\sqrt{2/7}\,\psi^{\alpha}_{43-1}$ | $\sqrt{2/21}\,\psi^{\alpha}_{430}$ | $-2\sqrt{2/21}\,\psi^{\alpha}_{431}$ | 0 | 0 | 0 | 0 |
| | | | -3/2 | 1 | 4 | $\sqrt{2/21}\,\psi^{\alpha}_{43-3}$ | $\sqrt{3/7}\,\psi^{\alpha}_{43-2}$ | 0 | $-\sqrt{10/21}\,\psi^{\alpha}_{430}$ | 0 | 0 | 0 | 0 |
| | | | -5/2 | 1 | 4 | 0 | $-2\sqrt{2/21}\,\psi^{\alpha}_{43-3}$ | $-1/\sqrt{7}\,\psi^{\alpha}_{43-2}$ | $\sqrt{10/21}\,\psi^{\alpha}_{43-1}$ | 0 | 0 | 0 | 0 |
| | | | -7/2 | 1 | 4 | 0 | 0 | $\sqrt{2/3}\,\psi^{\alpha}_{43-3}$ | $-1/\sqrt{3}\,\psi^{\alpha}_{43-2}$ | 0 | 0 | 0 | 0 |
| | | 9/2 | 9/2 | 3 | 6 | $\psi^{\alpha}_{433}$ | 0 | 0 | 0 | 0 | 0 | 0 | 0 |
| | | | 7/2 | 3 | 6 | $\sqrt{2/3}\,\psi^{\alpha}_{432}$ | $-1/\sqrt{3}\,\psi^{\alpha}_{433}$ | 0 | 0 | 0 | 0 | 0 | 0 |
| | | | 5/2 | 3 | 6 | $1/2\sqrt{5/3}\,\psi^{\alpha}_{431}$ | $-1/\sqrt{2}\,\psi^{\alpha}_{432}$ | $1/2\sqrt{3}\,\psi^{\alpha}_{433}$ | 0 | 0 | 0 | 0 | 0 |
| | | | 3/2 | 3 | 6 | $\sqrt{5/21}\,\psi^{\alpha}_{430}$ | $-1/2\sqrt{15/7}\,\psi^{\alpha}_{431}$ | $\sqrt{3/14}\,\psi^{\alpha}_{432}$ | $-1/2\sqrt{21}\,\psi^{\alpha}_{433}$ | 0 | 0 | 0 | 0 |
| | | | 1/2 | 3 | 6 | $-\sqrt{5/42}\,\psi^{\alpha}_{43-1}$ | $-\sqrt{10/21}\,\psi^{\alpha}_{430}$ | $\sqrt{5/14}\,\psi^{\alpha}_{431}$ | $-1/\sqrt{21}\,\psi^{\alpha}_{432}$ | 0 | 0 | 0 | 0 |
| | | | -1/2 | 3 | 6 | $1/\sqrt{21}\,\psi^{\alpha}_{43-2}$ | $\sqrt{5/14}\,\psi^{\alpha}_{43-1}$ | $\sqrt{10/21}\,\psi^{\alpha}_{430}$ | $-\sqrt{5/42}\,\psi^{\alpha}_{431}$ | 0 | 0 | 0 | 0 |
| | | | -3/2 | 3 | 6 | $-1/2\sqrt{21}\,\psi^{\alpha}_{43-3}$ | $-\sqrt{3/14}\,\psi^{\alpha}_{43-2}$ | $-1/2\sqrt{15/7}\,\psi^{\alpha}_{43-1}$ | $-\sqrt{5/21}\,\psi^{\alpha}_{430}$ | 0 | 0 | 0 | 0 |
| | | | -5/2 | 3 | 6 | 0 | $1/2\sqrt{3}\,\psi^{\alpha}_{43-3}$ | $1/\sqrt{2}\,\psi^{\alpha}_{43-2}$ | $1/2\sqrt{5/3}\,\psi^{\alpha}_{43-1}$ | 0 | 0 | 0 | 0 |
| | | | -7/2 | 3 | 6 | 0 | 0 | $-1/\sqrt{3}\,\psi^{\alpha}_{43-3}$ | $-\sqrt{2/3}\,\psi^{\alpha}_{43-2}$ | 0 | 0 | 0 | 0 |
| | | | -9/2 | 3 | 6 | 0 | 0 | 0 | $\psi^{\alpha}_{43-3}$ | 0 | 0 | 0 | 0 |

**Table 3.** The values of coefficients $_kA^{s\lambda}_{ljm}$, $_kB^{s\lambda}_{ljm}$, $_kC^{s\lambda}_{\tilde{l}jm}$ and $_kD^{s\lambda}_{\tilde{l}jm}$ for $s=1/2$, $0\le l\le 3$, $\frac{1}{2}\le j\le\frac{7}{2}$, $-j\le m\le j$ and $k=\pm 1$

| $l$ | $j$ | $m$ | $t$ | $\tilde{l}$ | $_1A^{\frac{1}{2}0}_{ljm}$ | $_{-1}A^{\frac{1}{2}0}_{ljm}$ | $_1B^{\frac{1}{2}0}_{ljm}$ | $_{-1}B^{\frac{1}{2}0}_{ljm}$ | $_1C^{\frac{1}{2}0}_{\tilde{l}jm}$ | $_{-1}C^{\frac{1}{2}0}_{\tilde{l}jm}$ | $_1D^{\frac{1}{2}0}_{\tilde{l}jm}$ | $_{-1}D^{\frac{1}{2}0}_{\tilde{l}jm}$ |
|---|---|---|---|---|---|---|---|---|---|---|---|---|
| 0 | 1/2 | 1/2 | 1 | 1 | $1/\sqrt{3}$ | 0 | $\sqrt{2/3}$ | 0 | $\sqrt{2/5}$ | 0 | $\sqrt{3/5}$ | 0 |
| | | -1/2 | 1 | 1 | $\sqrt{2/3}$ | 0 | $1/\sqrt{3}$ | 0 | $\sqrt{3/5}$ | 0 | $\sqrt{2/5}$ | 0 |
| 1 | 1/2 | 1/2 | -1 | 0 | 0 | -1 | 0 | 0 | 0 | 0 | 0 | -1 |
| | | -1/2 | -1 | 0 | 0 | 0 | 0 | 1 | 0 | 1 | 0 | 0 |
| | 3/2 | 3/2 | 1 | 2 | $1/\sqrt{5}$ | 0 | $2/\sqrt{5}$ | 0 | $\sqrt{2/7}$ | 0 | $\sqrt{5/7}$ | 0 |
| | | 1/2 | 1 | 2 | $\sqrt{2/5}$ | 0 | $\sqrt{3/5}$ | 0 | $\sqrt{3/7}$ | 0 | $2/\sqrt{7}$ | 0 |
| | | -1/2 | 1 | 2 | $\sqrt{3/5}$ | 0 | $\sqrt{2/5}$ | 0 | $2/\sqrt{7}$ | 0 | $\sqrt{3/7}$ | 0 |
| | | -3/2 | 1 | 2 | $2/\sqrt{5}$ | 0 | $1/\sqrt{5}$ | 0 | $\sqrt{5/7}$ | 0 | $\sqrt{2/7}$ | 0 |
| 2 | 3/2 | 3/2 | -1 | 1 | 0 | -1 | 0 | 0 | 0 | $\sqrt{2}$ | 0 | $-i$ |
| | | 1/2 | -1 | 1 | 0 | $-\sqrt{2/3}$ | 0 | $1/\sqrt{3}$ | 0 | 1 | 0 | 0 |
| | | -1/2 | -1 | 1 | 0 | $-1/\sqrt{3}$ | 0 | $\sqrt{2/3}$ | 0 | 0 | 0 | -1 |
| | | -3/2 | -1 | 1 | 0 | 0 | 0 | 1 | 0 | $i$ | 0 | $-\sqrt{2}$ |
| | 5/2 | 5/2 | 1 | 3 | $1/\sqrt{7}$ | 0 | $\sqrt{6/7}$ | 0 | $\sqrt{2}/3$ | 0 | $\sqrt{7}/3$ | 0 |
| | | 3/2 | 1 | 3 | $\sqrt{2/7}$ | 0 | $\sqrt{5/7}$ | 0 | $1/\sqrt{3}$ | 0 | $\sqrt{2/3}$ | 0 |
| | | 1/2 | 1 | 3 | $\sqrt{3/7}$ | 0 | $2/\sqrt{7}$ | 0 | $2/3$ | 0 | $\sqrt{5}/3$ | 0 |
| | | -1/2 | 1 | 3 | $2/\sqrt{7}$ | 0 | $\sqrt{3/7}$ | 0 | $\sqrt{5}/3$ | 0 | $2/3$ | 0 |
| | | -3/2 | 1 | 3 | $\sqrt{5/7}$ | 0 | $\sqrt{2/7}$ | 0 | $\sqrt{2/3}$ | 0 | $1/\sqrt{3}$ | 0 |
| | | -5/2 | 1 | 3 | $\sqrt{6/7}$ | 0 | $1/\sqrt{7}$ | 0 | $\sqrt{7}/3$ | 0 | $\sqrt{2/3}$ | 0 |
| 3 | 5/2 | 5/2 | -1 | 2 | 0 | -1 | 0 | 0 | 0 | $2/\sqrt{3}$ | 0 | $-i/\sqrt{3}$ |
| | | 3/2 | -1 | 2 | 0 | $-2/\sqrt{5}$ | 0 | $1/\sqrt{5}$ | 0 | 1 | 0 | 0 |
| | | 1/2 | -1 | 2 | 0 | $-\sqrt{3/5}$ | 0 | $\sqrt{2/5}$ | 0 | $\sqrt{2/3}$ | 0 | $-1/\sqrt{3}$ |
| | | -1/2 | -1 | 2 | 0 | $-\sqrt{2/5}$ | 0 | $\sqrt{3/5}$ | 0 | $1/\sqrt{3}$ | 0 | $-\sqrt{2/3}$ |
| | | -3/2 | -1 | 2 | 0 | $-1/\sqrt{5}$ | 0 | $2/\sqrt{5}$ | 0 | 0 | 0 | -1 |
| | | -5/2 | -1 | 2 | 0 | 0 | 0 | 1 | 0 | $i/\sqrt{3}$ | 0 | $-2/\sqrt{3}$ |
| | 7/2 | 7/2 | 1 | 4 | $1/3$ | 0 | $2\sqrt{2}/3$ | 0 | $\sqrt{2/11}$ | 0 | $3/\sqrt{11}$ | 0 |
| | | 5/2 | 1 | 4 | $\sqrt{2}/3$ | 0 | $\sqrt{7}/3$ | 0 | $\sqrt{3/11}$ | 0 | $2\sqrt{2/11}$ | 0 |
| | | 3/2 | 1 | 4 | $1/\sqrt{3}$ | 0 | $\sqrt{2/3}$ | 0 | $2/\sqrt{11}$ | 0 | $\sqrt{7/11}$ | 0 |
| | | 1/2 | 1 | 4 | $2/3$ | 0 | $\sqrt{5}/3$ | 0 | $\sqrt{5/11}$ | 0 | $\sqrt{6/11}$ | 0 |
| | | -1/2 | 1 | 4 | $\sqrt{5}/3$ | 0 | $2/3$ | 0 | $\sqrt{6/11}$ | 0 | $\sqrt{5/11}$ | 0 |
| | | -3/2 | 1 | 4 | $\sqrt{2/3}$ | 0 | $1/\sqrt{3}$ | 0 | $\sqrt{7/11}$ | 0 | $2/\sqrt{11}$ | 0 |
| | | -5/2 | 1 | 4 | $\sqrt{7}/3$ | 0 | $\sqrt{2}/3$ | 0 | $2\sqrt{2/11}$ | 0 | $\sqrt{3/11}$ | 0 |
| | | -7/2 | 1 | 4 | $2\sqrt{2}/3$ | 0 | $1/3$ | 0 | $3/\sqrt{11}$ | 0 | $\sqrt{2/11}$ | 0 |

**Table 4**. The values of coefficients ${}_kA^{s\lambda}_{ljm}$, ${}_kB^{s\lambda}_{ljm}$, ${}_kC^{s\lambda}_{\tilde{l}jm}$ and ${}_kD^{s\lambda}_{\tilde{l}jm}$ for $s=3/2$, $0 \le l \le 3$, $\frac{3}{2} \le j \le \frac{9}{2}$, $-j \le m \le j$ and $k=\pm 1$

| $\lambda$ | $l$ | $j$ | $m$ | $t$ | $\tilde{l}$ | ${}_1A^{\frac32\lambda}_{ljm}$ | ${}_{-1}A^{\frac32\lambda}_{ljm}$ | ${}_1B^{\frac32\lambda}_{ljm}$ | ${}_{-1}B^{\frac32\lambda}_{ljm}$ | ${}_1C^{\frac32\lambda}_{\tilde{l}jm}$ | ${}_{-1}C^{\frac32\lambda}_{\tilde{l}jm}$ | ${}_1D^{\frac32\lambda}_{\tilde{l}jm}$ | ${}_{-1}D^{\frac32\lambda}_{\tilde{l}jm}$ |
|---|---|---|---|---|---|---|---|---|---|---|---|---|---|
| 0 | 0 | 3/2 | 3/2 | 3 | 3 | $\frac{1}{\sqrt{3}}$ | 0 | $\sqrt{\frac{2}{3}}$ | 0 | $\frac{2}{21}(4+\sqrt{6})$ | $-\frac{2(-3+\sqrt{6})}{7\sqrt{5}}$ | $\frac{1}{21}\sqrt{5}(4+\sqrt{6})$ | $-\frac{2(-2+\sqrt{6})}{7\sqrt{5}}$ |
| | | | 1/2 | 3 | 3 | $\sqrt{\frac{2}{3}}$ | 0 | $\frac{1}{\sqrt{3}}$ | 0 | $\frac{4+\sqrt{10}}{7\sqrt{3}}$ | $\frac{2}{35}(-3\sqrt{2}+2\sqrt{5})$ | $\frac{2}{105}(5\sqrt{6}+4\sqrt{15})$ | $-\frac{2}{35}\sqrt{3}(-3+\sqrt{10})$ |
| | | | -1/2 | 3 | 3 | 0 | 0 | 0 | 0 | $\frac{\sqrt{6}}{7}+\frac{4}{7\sqrt{15}}$ | $\frac{1}{35}(-3\sqrt{2}+2\sqrt{5})$ | $\frac{15+2\sqrt{10}}{35\sqrt{3}}$ | $\frac{2}{35}(3\sqrt{2}-2\sqrt{5})$ |
| | | | -3/2 | 3 | 3 | 0 | 0 | 0 | 0 | $\frac{1+2\sqrt{6}}{3\sqrt{35}}$ | 0 | $\frac{1}{105}(\sqrt{10}+4\sqrt{15})$ | $-\frac{-2+\sqrt{6}}{7\sqrt{5}}$ |
| | 1 | 3/2 | 3/2 | 1 | 2 | $\frac{2}{15}(-3+\sqrt{3})$ | $-\frac{3+2\sqrt{3}}{3\sqrt{5}}$ | $\frac{1}{5}(\sqrt{2}-\sqrt{6})$ | 0 | $-\frac{2\sqrt{\frac{3}{7}}}{5}$ | $\frac{2}{5}$ | $-\frac{4}{5\sqrt{7}}$ | $-\frac{\sqrt{2}}{5}$ |
| | | | 1/2 | 1 | 2 | $-\frac{1}{15}\sqrt{2}(3+\sqrt{3})$ | 0 | $-\frac{2}{15}(1+\sqrt{3})$ | $\frac{-1+2\sqrt{3}}{3\sqrt{5}}$ | $\frac{4}{5\sqrt{7}}$ | $\frac{\sqrt{2}}{5}$ | $\frac{2\sqrt{\frac{3}{7}}}{5}$ | $-\frac{2}{5}$ |
| | | | -1/2 | 1 | 2 | $-\frac{4\sqrt{\frac{2}{3}}}{5}$ | 0 | $-\frac{2\sqrt{\frac{2}{3}}}{5}$ | 0 | $\frac{2+\sqrt{2}}{\sqrt{35}}$ | 0 | $\frac{2}{35}(\sqrt{7}+\sqrt{14})$ | $\frac{1}{5}(1-2\sqrt{2})$ |
| | | | -3/2 | 1 | 2 | 0 | 0 | 0 | 0 | $2\sqrt{\frac{3}{35}}$ | $\frac{i}{\sqrt{15}}$ | $\sqrt{\frac{2}{35}}$ | $-\frac{2}{\sqrt{15}}$ |
| | | 5/2 | 5/2 | 3 | 4 | $\frac{1}{\sqrt{5}}$ | 0 | $\frac{2}{\sqrt{5}}$ | 0 | $\frac{4+\sqrt{3}}{3\sqrt{11}}$ | $-\frac{1}{3}\sqrt{\frac{5}{14}}(-2+\sqrt{3})$ | $\frac{1}{6}\sqrt{\frac{7}{11}}(4+\sqrt{3})$ | $\frac{-2+\sqrt{3}}{3\sqrt{7}}$ |
| | | | 3/2 | 3 | 4 | $\frac{1}{15}\sqrt{2}(3+2\sqrt{3})$ | $\frac{1}{3}\sqrt{\frac{2}{5}}(-3+\sqrt{3})$ | $\frac{1}{5}(2+\sqrt{3})$ | 0 | $\frac{5\sqrt{\frac{2}{77}}}{3}+\frac{5\sqrt{\frac{5}{33}}}{6}$ | $\frac{1}{63}(-15\sqrt{2}+2\sqrt{105})$ | $2\sqrt{\frac{5}{231}}+\frac{5}{3\sqrt{22}}$ | $\frac{1}{42}(5\sqrt{6}-2\sqrt{35})$ |
| | | | ½ | 3 | 4 | $\frac{1}{10}(\sqrt{2}+2\sqrt{6})$ | 0 | $\frac{1}{15}(6+\sqrt{3})$ | $\frac{1}{\sqrt{5}}-\frac{1}{\sqrt{15}}$ | $\frac{2\sqrt{\frac{10}{77}}}{3}+\frac{5}{\sqrt{231}}$ | $\frac{1}{21}(5-\sqrt{30})$ | $\frac{5(4\sqrt{231}+3\sqrt{770})}{1386}$ | $-\frac{2}{63}(5\sqrt{3}-3\sqrt{10})$ |
| | | | -1/2 | 3 | 4 | $\frac{\sqrt{6}}{5}$ | 0 | $\frac{\sqrt{\frac{3}{2}}}{5}$ | 0 | $\frac{1}{66}(4\sqrt{33}+\sqrt{110})$ | $\frac{1}{21}(-2\sqrt{3}+\sqrt{10})$ | $\frac{1}{231}(4\sqrt{231}+\sqrt{770})$ | $\frac{1}{21}(-5+\sqrt{30})$ |
| | | | -3/2 | 3 | 4 | 0 | 0 | 0 | 0 | $\frac{1}{693}(21\sqrt{110}+4\sqrt{231})$ | $\frac{1}{126}(-3\sqrt{10}+2\sqrt{21})$ | $\frac{1}{462}(2\sqrt{154}+7\sqrt{165})$ | $\frac{1}{21}(-\sqrt{14}+\sqrt{15})$ |
| | | | -5/2 | 3 | 4 | 0 | 0 | 0 | 0 | $\frac{12+\sqrt{3}}{6\sqrt{77}}$ | 0 | $\frac{12+\sqrt{3}}{9\sqrt{154}}$ | $\frac{3-2\sqrt{3}}{9\sqrt{14}}$ |
| | 2 | 3/2 | 3/2 | -1 | 1 | $\frac{3-2\sqrt{3}}{5\sqrt{7}}$ | $\frac{2}{15}(3+\sqrt{3})$ | $\frac{2(-2+\sqrt{3})}{5\sqrt{7}}$ | $-\frac{1}{15}\sqrt{2}(3+\sqrt{3})$ | 0 | 0 | 0 | 0 |
| | | | 1/2 | -1 | 1 | $-\frac{2(-2+\sqrt{3})}{5\sqrt{7}}$ | $\frac{1}{15}\sqrt{2}(3+\sqrt{3})$ | $\frac{-3+2\sqrt{3}}{5\sqrt{7}}$ | $-\frac{2}{15}(3+\sqrt{3})$ | $-\frac{4}{5\sqrt{3}}$ | 0 | $-\frac{4\sqrt{2}}{15}$ | $-\frac{2\sqrt{\frac{2}{5}}}{3}$ |
| | | | -1/2 | -1 | 1 | $\sqrt{\frac{2}{35}}$ | 0 | $\frac{2}{5\sqrt{7}}$ | $-\frac{2\sqrt{2}}{5}$ | $-\frac{2}{5\sqrt{3}}$ | $i\sqrt{\frac{2}{5}}$ | $-\frac{1}{5\sqrt{3}}$ | $-\frac{2}{\sqrt{5}}$ |
| | | | -3/2 | -1 | 1 | $2\sqrt{\frac{3}{35}}$ | 0 | $\sqrt{\frac{2}{35}}$ | 0 | $\frac{3i+2\sqrt{6}}{3\sqrt{5}}$ | $2i\sqrt{\frac{2}{5}-\frac{2}{\sqrt{15}}}$ | 0 | $i\sqrt{\frac{2}{5}-2\sqrt{\frac{3}{5}}}$ |
| | | 5/2 | 5/2 | 1 | 3 | $-\frac{2}{7}\sqrt{\frac{2}{5}}(-1+\sqrt{3})$ | $-\frac{4+\sqrt{3}}{\sqrt{35}}$ | $-\frac{2}{7}(-1+\sqrt{3})$ | 0 | $-\frac{\sqrt{\frac{5}{2}}}{7}$ | $\frac{\sqrt{15}}{7}$ | $-\frac{\sqrt{5}}{7}$ | $-\frac{\sqrt{15}}{14}$ |
| | | | 3/2 | 1 | 3 | $\frac{1}{35}\sqrt{2}(-9+\sqrt{3})$ | $-\frac{1}{5}\sqrt{\frac{2}{7}}(1+2\sqrt{3})$ | $\frac{2}{35}\sqrt{2}(1-3\sqrt{3})$ | $\frac{1}{35}(\sqrt{7}+2\sqrt{21})$ | $\frac{1}{21}(6-\sqrt{6})$ | $\frac{9+2\sqrt{6}}{14\sqrt{5}}$ | $-\frac{1}{42}\sqrt{5}(-6+\sqrt{6})$ | $-\frac{4+3\sqrt{6}}{14\sqrt{5}}$ |
| | | | ½ | 1 | 3 | $-\frac{6}{35}(1+\sqrt{3})$ | $\frac{2-3\sqrt{3}}{5\sqrt{14}}$ | $-\frac{3}{35}(3+\sqrt{3})$ | $\frac{-2+3\sqrt{3}}{5\sqrt{7}}$ | $\frac{1}{21}(\sqrt{2}+3\sqrt{5})$ | $\frac{1}{35}\sqrt{3}(-1+2\sqrt{10})$ | $\frac{2}{105}(15+\sqrt{10})$ | $\frac{3}{70}(\sqrt{2}-4\sqrt{5})$ |
| | | | -1/2 | 1 | 3 | $-\frac{1}{7}\sqrt{\frac{2}{5}}(5+\sqrt{3})$ | 0 | $-\frac{2}{35}(5+\sqrt{3})$ | $\frac{-5+4\sqrt{3}}{5\sqrt{14}}$ | $\frac{2}{7}+\frac{3\sqrt{\frac{2}{5}}}{7}$ | $\frac{1}{70}\sqrt{3}(-4+3\sqrt{10})$ | $\frac{1}{35}(5\sqrt{2}+3\sqrt{5})$ | $\frac{1}{35}(4\sqrt{3}-3\sqrt{30})$ |
| | | | -3/2 | 1 | 3 | $-\frac{4\sqrt{\frac{6}{5}}}{7}$ | 0 | $-\frac{4}{7\sqrt{5}}$ | 0 | $\sqrt{\frac{7}{30}+\frac{1}{\sqrt{35}}}$ | 0 | $\frac{\sqrt{\frac{2}{5}}}{7}+\frac{1}{\sqrt{15}}$ | $\frac{7-6\sqrt{6}}{14\sqrt{5}}$ |
| | | | -5/2 | 1 | 3 | 0 | 0 | 0 | 0 | $\frac{2\sqrt{\frac{5}{7}}}{3}$ | $\frac{1}{2}i\sqrt{\frac{3}{35}}$ | $\frac{\sqrt{\frac{5}{14}}}{3}$ | $-\frac{3}{\sqrt{70}}$ |
| | | 7/2 | 7/2 | 3 | 5 | 1 | 0 | $\sqrt{\frac{6}{7}}$ | 0 | $\frac{4(20+3\sqrt{5})}{55\sqrt{13}}$ | $-\frac{2}{55}\sqrt{\frac{14}{3}}(-5+2\sqrt{5})$ | $\frac{6(20+3\sqrt{5})}{55\sqrt{13}}$ | $\frac{4(-5+2\sqrt{5})}{55\sqrt{3}}$ |
| | | | 5/2 | 3 | 5 | $\frac{1}{7}\sqrt{\frac{2}{5}}(4+\sqrt{3})$ | $-\frac{2(-1+\sqrt{3})}{\sqrt{35}}$ | $\frac{1}{7}(4+\sqrt{3})$ | 0 | $\frac{2(21+2\sqrt{42})}{33\sqrt{13}}$ | $\frac{4(-7+\sqrt{42})}{33\sqrt{5}}$ | $\frac{4(21\sqrt{130}+4\sqrt{1365})}{2145}$ | $\frac{2}{165}(7\sqrt{10}-2\sqrt{105})$ |
| | | | 3/2 | 3 | 5 | $\frac{1}{7}\sqrt{\frac{2}{5}}(3+2\sqrt{3})$ | $\frac{2}{3}\sqrt{\frac{2}{35}}(-3+\sqrt{3})$ | $\frac{2}{7}\sqrt{\frac{2}{5}}(2+\sqrt{3})$ | $-\frac{2(-3+\sqrt{3})}{3\sqrt{35}}$ | $\frac{1}{55}\sqrt{\frac{14}{13}}(15+4\sqrt{15})$ | $\frac{1}{99}\sqrt{14}(-9+2\sqrt{15})$ | $\frac{7}{715}(5\sqrt{39}+4\sqrt{65})$ | $\frac{2}{495}(-10\sqrt{42}+9\sqrt{70})$ |
| | | | 1/2 | 3 | 5 | $\frac{2}{35}\sqrt{2}(2+3\sqrt{3})$ | $-\frac{2(-1+\sqrt{3})}{5\sqrt{7}}$ | $\frac{1}{35}\sqrt{2}(9+2\sqrt{3})$ | $\frac{2}{5}\sqrt{\frac{2}{7}}(-1+\sqrt{3})$ | $\frac{1}{429}(7\sqrt{195}+6\sqrt{546})$ | $\frac{2}{33}(-2\sqrt{5}+\sqrt{14})$ | $\frac{12+\sqrt{70}}{11\sqrt{13}}$ | $\frac{1}{33}(10-\sqrt{70})$ |
| | | | -1/2 | 3 | 5 | $\frac{1}{35}(\sqrt{5}+4\sqrt{15})$ | 0 | $\frac{1}{35}(\sqrt{2}+4\sqrt{6})$ | $\frac{2(-1+\sqrt{3})}{5\sqrt{7}}$ | $\frac{4}{429}(2\sqrt{78}+\sqrt{1365})$ | $\frac{2}{33}(\sqrt{7}-\sqrt{10})$ | $\frac{1}{429}(4\sqrt{195}+5\sqrt{546})$ | $\frac{2}{33}(2\sqrt{5}-\sqrt{14})$ |
| | | | -3/2 | 3 | 5 | $\frac{\sqrt{6}}{7}$ | 0 | $\frac{1}{7}$ | 0 | $\frac{3}{55}\sqrt{\frac{2}{13}}(5+4\sqrt{15})$ | $\frac{2}{495}(10\sqrt{6}-9\sqrt{10})$ | $\frac{2}{55}\sqrt{\frac{2}{13}}(5+4\sqrt{15})$ | $\frac{2}{495}(-10\sqrt{21}+9\sqrt{35})$ |
| | | | -5/2 | 3 | 5 | 0 | 0 | 0 | 0 | $\frac{1}{429}(4\sqrt{39}+9\sqrt{182})$ | $\frac{1}{165}(2\sqrt{15}-\sqrt{70})$ | $\frac{1}{715}(2\sqrt{130}+3\sqrt{1365})$ | $-\frac{4}{165}(\sqrt{30}-\sqrt{35})$ |
| | | | -7/2 | 3 | 5 | 0 | 0 | 0 | 0 | $\frac{20+\sqrt{5}}{5\sqrt{429}}$ | 0 | $\frac{1}{55}\sqrt{\frac{2}{39}}(20+\sqrt{5})$ | $\frac{1}{55}\sqrt{\frac{2}{3}}(-5+2\sqrt{5})$ |
| | 3 | 3/2 | 3/2 | -3 | 0 | $\frac{4(-1+\sqrt{3})}{21\sqrt{5}}$ | $\frac{1}{35}(-3-4\sqrt{3})$ | $\frac{2}{21}(-1+\sqrt{3})$ | $\frac{1}{35}\sqrt{2}(4+\sqrt{3})$ | 0 | 0 | 0 | 0 |
| | | | 1/2 | -3 | 0 | $-\frac{2}{21}(-3+\sqrt{3})$ | $-\frac{1}{35}\sqrt{2}(4+3\sqrt{3})$ | $-\frac{4(-3+\sqrt{3})}{21\sqrt{5}}$ | $\frac{1}{35}(9+4\sqrt{3})$ | 0 | 0 | 0 | 0 |
| | | | -1/2 | -3 | 0 | $-\frac{2}{21}\sqrt{2}(-3+\sqrt{3})$ | $-\frac{1}{35}\sqrt{2}(5+2\sqrt{3})$ | $-\frac{2}{21}(-3+\sqrt{3})$ | $\frac{2}{35}\sqrt{2}(5+2\sqrt{3})$ | $-\sqrt{2}$ | $i\sqrt{2}$ | 0 | $-1$ |
| | | | -3/2 | -3 | 0 | $\frac{2(-1+\sqrt{3})}{3\sqrt{7}}$ | 0 | $\frac{2}{21}\sqrt{2}(-1+\sqrt{3})$ | $\frac{1}{7}(\sqrt{2}+2\sqrt{6})$ | $2i\sqrt{\frac{2}{3}}$ | 3 | $\sqrt{\frac{2}{3}}$ | $-i\sqrt{6}$ |
| | | 5/2 | 5/2 | -1 | 2 | $\frac{1}{42}\sqrt{5}(3-2\sqrt{3})$ | $\frac{1}{7}\sqrt{\frac{2}{5}}(5+\sqrt{3})$ | $\frac{1}{21}\sqrt{\frac{5}{2}}(3-2\sqrt{3})$ | $-\frac{5+\sqrt{3}}{7\sqrt{10}}$ | 0 | 0 | 0 | 0 |
| | | | 3/2 | -1 | 2 | $\frac{4}{7\sqrt{5}}-\frac{1}{\sqrt{15}}$ | $\frac{1}{35}(9+7\sqrt{3})$ | $\frac{2}{7}-\frac{1}{2\sqrt{3}}$ | $-\frac{1}{35}\sqrt{2}(7+3\sqrt{3})$ | $-\frac{4\sqrt{6}}{35}$ | $\frac{4\sqrt{\frac{2}{7}}}{5}$ | $-\frac{8\sqrt{2}}{35}$ | $-\frac{4}{5\sqrt{7}}$ |
| | | | 1/2 | -1 | 2 | $\frac{1}{42}(9\sqrt{2}-4\sqrt{6})$ | $\frac{6}{35}(1+\sqrt{3})$ | $\frac{1}{21}\sqrt{\frac{2}{5}}(9-4\sqrt{3})$ | $-\frac{3}{35}\sqrt{2}(3+\sqrt{3})$ | $\frac{1}{35}\sqrt{3}(4-5\sqrt{2})$ | $\frac{\sqrt{\frac{6}{7}}}{5}+\frac{1}{\sqrt{21}}$ | $\frac{3}{70}(4-5\sqrt{2})$ | $-\frac{6+5\sqrt{2}}{5\sqrt{21}}$ |
| | | | -1/2 | -1 | 2 | $\frac{1}{21}(6-\sqrt{3})$ | $\frac{1}{35}(1+5\sqrt{3})$ | $-\frac{-6+\sqrt{3}}{21\sqrt{2}}$ | $-\frac{2}{35}(1+5\sqrt{3})$ | $\frac{1}{14}\sqrt{\frac{3}{5}}(-4+3\sqrt{2})$ | 0 | $\frac{1}{35}(3\sqrt{3}-2\sqrt{6})$ | $-\frac{1}{5}\sqrt{\frac{3}{7}}(1+3\sqrt{2})$ |
| | | | -3/2 | -1 | 2 | $\frac{1}{2\sqrt{7}}+\frac{1}{\sqrt{21}}$ | 0 | $\frac{3+2\sqrt{3}}{21\sqrt{2}}$ | $\frac{1}{14}(\sqrt{2}-3\sqrt{6})$ | $\frac{\sqrt{\frac{6}{5}}}{7}$ | $i\sqrt{\frac{6}{35}}$ | $\frac{1}{7\sqrt{5}}$ | $-2\sqrt{\frac{6}{35}}$ |
| | | | -5/2 | -1 | 2 | $\sqrt{\frac{10}{21}}$ | 0 | $\frac{\sqrt{\frac{5}{21}}}{2}$ | 0 | $\sqrt{\frac{3}{35}}(i+2\sqrt{2})$ | $-\frac{2(-3i+\sqrt{2})}{\sqrt{105}}$ | 0 | $\frac{2i-3\sqrt{2}}{\sqrt{21}}$ |
| | | 7/2 | 7/2 | 1 | 4 | $\frac{2(-3+\sqrt{3})}{9\sqrt{7}}$ | $-\frac{6+\sqrt{3}}{3\sqrt{7}}$ | $\frac{1}{9}\sqrt{2}(-3+\sqrt{3})$ | 0 | $-\frac{2\sqrt{\frac{14}{165}}}{3}$ | $\frac{2\sqrt{\frac{14}{15}}}{3}$ | $-\frac{8\sqrt{\frac{7}{55}}}{9}$ | $-\frac{2\sqrt{\frac{7}{5}}}{9}$ |
| | | | 5/2 | 1 | 4 | $\frac{1}{21}\sqrt{2}(-5+\sqrt{3})$ | $-\frac{2}{21}(3+2\sqrt{3})$ | $\frac{2}{21}(-5+\sqrt{3})$ | $\frac{1}{21}(3+2\sqrt{3})$ | $-\frac{4}{9}\sqrt{\frac{2}{55}}(-3+\sqrt{3})$ | $\frac{2(3+2\sqrt{3})}{9\sqrt{7}}$ | $-\frac{2}{9}\sqrt{\frac{14}{55}}(-3+\sqrt{3})$ | $-\frac{2}{9}\sqrt{\frac{2}{35}}(3+2\sqrt{3})$ |
| | | | 3/2 | 1 | 4 | $-\frac{4\sqrt{\frac{10}{3}}}{21}$ | $-\frac{\sqrt{6}}{7}$ | $-\frac{10\sqrt{\frac{2}{3}}}{21}$ | $\frac{2}{7}$ | $\frac{1}{693}(21\sqrt{110}-2\sqrt{231})$ | $\frac{2}{315}(5\sqrt{3}+6\sqrt{70})$ | $-\frac{2\sqrt{\frac{2}{385}}}{3}+\frac{2}{\sqrt{33}}$ | $-\frac{1}{21}-2\sqrt{\frac{2}{105}}$ |
| | | | 1/2 | 1 | 4 | $-\frac{2}{63}\sqrt{10}(3+\sqrt{3})$ | $\frac{2(3-4\sqrt{3})}{21\sqrt{5}}$ | $-\frac{4}{63}\sqrt{2}(3+\sqrt{3})$ | $-\frac{1}{7}\sqrt{\frac{2}{5}}(-4+\sqrt{3})$ | $\frac{2}{693}(9\sqrt{154}+\sqrt{1155})$ | $\frac{1}{63}(9\sqrt{5}-2\sqrt{6})$ | $2\sqrt{\frac{5}{231}}+\frac{5\sqrt{\frac{2}{77}}}{9}$ | $\frac{2}{63}(2\sqrt{2}-3\sqrt{15})$ |
| | | | -1/2 | 1 | 4 | $-\frac{2}{21}(\sqrt{5}+\sqrt{15})$ | $\frac{6-5\sqrt{3}}{21\sqrt{5}}$ | $-\frac{1}{21}\sqrt{10}(1+\sqrt{3})$ | $\frac{2(-6+5\sqrt{3})}{21\sqrt{5}}$ | $\frac{2}{693}(9\sqrt{154}+\sqrt{1155})$ | $-\frac{2}{63}(-6\sqrt{2}+\sqrt{15})$ | $\frac{4}{693}(3\sqrt{154}+\sqrt{1155})$ | $\frac{1}{63}(-12\sqrt{5}+5\sqrt{6})$ |
| | | | -3/2 | 1 | 4 | $-\frac{1}{9}\sqrt{\frac{2}{7}}(9+\sqrt{3})$ | 0 | $-\frac{2}{63}(9+\sqrt{3})$ | $\frac{1}{7}(2-\sqrt{3})$ | $\frac{4\sqrt{\frac{2}{77}}}{3}+\frac{16}{3\sqrt{165}}$ | $\frac{1}{315}(15\sqrt{14}-8\sqrt{15})$ | $\frac{4\sqrt{\frac{2}{55}}}{3}+\frac{2}{\sqrt{231}}$ | $\frac{8\sqrt{\frac{2}{5}}}{21}-\frac{2}{\sqrt{21}}$ |
| | | | -5/2 | 1 | 4 | $-\frac{8}{3\sqrt{21}}$ | 0 | $-\frac{2\sqrt{\frac{2}{21}}}{3}$ | 0 | $\frac{1}{3}\sqrt{\frac{2}{385}}(22+3\sqrt{3})$ | 0 | $\frac{2(22+3\sqrt{3})}{9\sqrt{385}}$ | $\frac{11-12\sqrt{3}}{9\sqrt{35}}$ |
| | | | -7/2 | 1 | 4 | 0 | 0 | 0 | 0 | $\frac{2\sqrt{\frac{7}{11}}}{3}$ | $\frac{i}{\sqrt{105}}$ | $\frac{\sqrt{\frac{14}{55}}}{3}$ | $-2\sqrt{\frac{2}{105}}$ |
| | | 9/2 | 9/2 | 3 | 6 | $\frac{1}{3}$ | 0 | $\frac{2\sqrt{2}}{3}$ | 0 | $\frac{1}{39}(3\sqrt{2}+4\sqrt{15})$ | $\frac{3(-5+\sqrt{30})}{13\sqrt{11}}$ | $\frac{1}{78}(3\sqrt{22}+4\sqrt{165})$ | $\frac{1}{143}(5\sqrt{22}-2\sqrt{165})$ |
| | | | 7/2 | 3 | 6 | $\frac{1}{9}\sqrt{\frac{2}{7}}(6+\sqrt{3}$ | $\frac{1}{3}\sqrt{\frac{2}{7}}(-3+\sqrt{3})$ | $\frac{1}{9}(6+\sqrt{3})$ | 0 | $\frac{1}{286}(33\sqrt{10}+4\sqrt{165})$ | $\frac{2}{143}(-9\sqrt{10}+2\sqrt{165})$ | $\frac{1}{143}\sqrt{5}(33+2\sqrt{66})$ | $-\frac{3}{143}(-3\sqrt{15}+\sqrt{110})$ |
| | | | 5/2 | 3 | 6 | $\frac{5}{6\sqrt{7}}+\frac{1}{\sqrt{21}}$ | $\frac{1}{3}\sqrt{\frac{2}{7}}(-3+\sqrt{3})$ | $\frac{5+2\sqrt{3}}{3\sqrt{14}}$ | $\frac{1}{\sqrt{14}}-\frac{1}{\sqrt{42}}$ | $\frac{14\sqrt{\frac{3}{11}}}{13}$ | $-\frac{2\sqrt{21}}{143}$ | $\frac{21\sqrt{\frac{2}{11}}}{13}$ | $\frac{4\sqrt{3}}{143}$ |
| | | | 3/2 | 3 | 6 | $\frac{1}{63}\sqrt{5}(9+4\sqrt{3})$ | $\frac{1}{7}(-3+\sqrt{3})$ | $\frac{5}{126}(9+4\sqrt{3})$ | $\frac{1}{7}\sqrt{2}(-1+\sqrt{3})$ | $\frac{7(4+\sqrt{14})}{13\sqrt{55}}$ | $\frac{2}{143}\sqrt{3}(-14+3\sqrt{14})$ | $\frac{4(7+2\sqrt{14})}{13\sqrt{55}}$ | $\frac{2}{143}(7\sqrt{10}-3\sqrt{35})$ |
| | | | 1/2 | 3 | 6 | $\frac{5(3+4\sqrt{3})}{63\sqrt{2}}$ | $\frac{2}{21}(-3+\sqrt{3})$ | $\frac{1}{63}\sqrt{10}(3+4\sqrt{3})$ | $\frac{1}{7}\sqrt{2}(-1+\sqrt{3})$ | $\frac{2(35+4\sqrt{35})}{65\sqrt{11}}$ | $\frac{1}{143}(14\sqrt{5}-15\sqrt{7})$ | $\frac{7(4\sqrt{110}+5\sqrt{154})}{1430}$ | $\frac{1}{143}\sqrt{6}(-14+3\sqrt{35})$ |
| | | | -1/2 | 3 | 6 | $\frac{1}{21}(2+5\sqrt{3})$ | $\frac{1}{21}(-3+\sqrt{3})$ | $\frac{2+5\sqrt{3}}{21\sqrt{2}}$ | $-\frac{2}{21}(-3+\sqrt{3})$ | $\frac{1}{13}\sqrt{\frac{3}{22}}(12+\sqrt{35})$ | $\frac{2}{143}(-15+2\sqrt{35})$ | $\frac{12+\sqrt{35}}{13\sqrt{11}}$ | $\frac{1}{143}(-14\sqrt{5}+15\sqrt{7})$ |
| | | | -3/2 | 3 | 6 | $\frac{1}{\sqrt{7}}+\frac{1}{6\sqrt{21}}$ | 0 | $\frac{18+\sqrt{3}}{63\sqrt{2}}$ | $\frac{-1+\sqrt{3}}{7\sqrt{2}}$ | $\frac{1}{13}\sqrt{\frac{5}{33}}(4+3\sqrt{14})$ | $\frac{1}{143}(3\sqrt{30}-2\sqrt{105})$ | $\frac{1}{429}(2\sqrt{330}+3\sqrt{1155})$ | $\frac{4}{143}\sqrt{5}(-3+\sqrt{14})$ |
| | | | -5/2 | 3 | 6 | $\frac{\sqrt{\frac{2}{3}}}{3}$ | 0 | $\frac{1}{6\sqrt{3}}$ | 0 | $\frac{5}{13}$ | $-\frac{2}{143}$ | $\frac{10}{13\sqrt{11}}$ | $\frac{3\sqrt{2}}{143}$ |
| | | | -7/2 | 3 | 6 | 0 | 0 | 0 | 0 | $\frac{1}{715}(33\sqrt{30}+4\sqrt{55})$ | $\frac{1}{143}(-3\sqrt{3}+\sqrt{22})$ | $\frac{33\sqrt{30}+4\sqrt{55}}{1430}$ | $\frac{1}{143}(3\sqrt{30}-2\sqrt{55})$ |
| | | | -9/2 | 3 | 6 | 0 | 0 | 0 | 0 | $\frac{2}{\sqrt{143}}+\frac{1}{\sqrt{4290}}$ | 0 | $\frac{30\sqrt{22}+\sqrt{165}}{2145}$ | $\frac{1}{143}(\sqrt{55}-\sqrt{66})$ |
| 2 | 0 | 3/2 | 3/2 | 3 | 3 | 0 | 0 | 0 | 0 | $\frac{1}{105}(\sqrt{10}+4\sqrt{15})$ | $-\frac{-2+\sqrt{6}}{7\sqrt{5}}$ | $\frac{1+2\sqrt{6}}{3\sqrt{35}}$ | 0 |
| | | | 1/2 | 3 | 3 | 0 | 0 | 0 | 0 | $\frac{15+2\sqrt{10}}{35\sqrt{3}}$ | $\frac{2}{35}(3\sqrt{2}-2\sqrt{5})$ | $\frac{\sqrt{6}}{7}+\frac{4}{7\sqrt{15}}$ | $\frac{1}{35}(-3\sqrt{2}+2\sqrt{5})$ |
| | | | -1/2 | 3 | 3 | $\frac{1}{\sqrt{3}}$ | 0 | $\sqrt{\frac{2}{3}}$ | 0 | $\frac{2}{105}(5\sqrt{6}+4\sqrt{15})$ | $-\frac{2}{35}\sqrt{3}(-3+\sqrt{10})$ | $\frac{4+\sqrt{10}}{7\sqrt{3}}$ | $\frac{2}{35}(-3\sqrt{2}+2\sqrt{5})$ |
| | | | -3/2 | 3 | 3 | $\sqrt{\frac{2}{3}}$ | 0 | $\frac{1}{\sqrt{3}}$ | 0 | $\frac{1}{21}\sqrt{5}(4+\sqrt{6})$ | $-\frac{2(-2+\sqrt{6})}{7\sqrt{5}}$ | $\frac{2}{21}(4+\sqrt{6})$ | $-\frac{2(-3+\sqrt{6})}{7\sqrt{5}}$ |
| | 1 | 3/2 | 3/2 | 1 | 2 | 0 | 0 | 0 | 0 | $-\sqrt{\frac{2}{35}}$ | $\frac{2}{\sqrt{15}}$ | $-2\sqrt{\frac{3}{35}}$ | $-\frac{i}{\sqrt{15}}$ |
| | | | 1/2 | 1 | 2 | $\frac{2\sqrt{\frac{2}{3}}}{5}$ | 0 | $\frac{4\sqrt{\frac{2}{3}}}{5}$ | 0 | $-\frac{2}{35}(\sqrt{7}+\sqrt{14})$ | $\frac{1}{5}(-1+2\sqrt{2})$ | $-\frac{2+\sqrt{2}}{\sqrt{35}}$ | 0 |
| | | | -1/2 | 1 | 2 | $\frac{2}{15}(1+\sqrt{3})$ | $\frac{1}{15}(\sqrt{5}-2\sqrt{15})$ | $\frac{1}{15}\sqrt{2}(3+\sqrt{3})$ | 0 | $-\frac{2\sqrt{\frac{3}{7}}}{5}$ | $\frac{2}{5}$ | $-\frac{4}{5\sqrt{7}}$ | $-\frac{\sqrt{2}}{5}$ |
| | | | -3/2 | 1 | 2 | $\frac{1}{5}\sqrt{2}(-1+\sqrt{3})$ | 0 | $-\frac{2}{15}(-3+\sqrt{3})$ | $\frac{1}{\sqrt{5}}+\frac{2}{\sqrt{15}}$ | $\frac{4}{5\sqrt{7}}$ | $\frac{\sqrt{2}}{5}$ | $\frac{2\sqrt{\frac{3}{7}}}{5}$ | $-\frac{2}{5}$ |
| | | 5/2 | 5/2 | 3 | 4 | 0 | 0 | 0 | 0 | $\frac{12+\sqrt{3}}{9\sqrt{154}}$ | $\frac{3-2\sqrt{3}}{9\sqrt{14}}$ | $\frac{12+\sqrt{3}}{6\sqrt{77}}$ | 0 |
| | | | 3/2 | 3 | 4 | 0 | 0 | 0 | 0 | $\frac{1}{462}(2\sqrt{154}+7\sqrt{165})$ | $\frac{1}{21}(-\sqrt{14}+\sqrt{15})$ | $\frac{1}{693}(21\sqrt{110}+4\sqrt{231})$ | $\frac{1}{126}(-3\sqrt{10}+2\sqrt{21})$ |
| | | | ½ | 3 | 4 | $\frac{\sqrt{\frac{3}{2}}}{5}$ | 0 | $\frac{\sqrt{6}}{5}$ | 0 | $\frac{1}{231}(4\sqrt{231}+\sqrt{770})$ | $\frac{1}{21}(-5+\sqrt{30})$ | $\frac{1}{66}(4\sqrt{33}+\sqrt{110})$ | $\frac{1}{21}(-2\sqrt{3}+\sqrt{10})$ |
| | | | -1/2 | 3 | 4 | $\frac{1}{15}(6+\sqrt{3})$ | $\frac{1}{\sqrt{5}}-\frac{1}{\sqrt{15}}$ | $\frac{1}{10}(\sqrt{2}+2\sqrt{6})$ | 0 | $\frac{5(4\sqrt{231}+3\sqrt{770})}{1386}$ | $-\frac{2}{63}(5\sqrt{3}-3\sqrt{10})$ | $\frac{2\sqrt{\frac{10}{77}}}{3}+\frac{5}{\sqrt{231}}$ | $\frac{1}{21}(5-\sqrt{30})$ |
| | | | -3/2 | 3 | 4 | $\frac{1}{5}(2+\sqrt{3})$ | 0 | $\frac{1}{15}\sqrt{2}(3+2\sqrt{3})$ | $\frac{1}{3}\sqrt{\frac{2}{5}}(-3+\sqrt{3})$ | $2\sqrt{\frac{5}{231}}+\frac{5}{3\sqrt{22}}$ | $\frac{1}{42}(5\sqrt{6}-2\sqrt{35})$ | $\frac{5\sqrt{\frac{2}{77}}}{3}+\frac{5\sqrt{\frac{5}{33}}}{6}$ | $\frac{1}{63}(-15\sqrt{2}+2\sqrt{105})$ |
| | | | -5/2 | 3 | 4 | $\frac{2}{\sqrt{5}}$ | 0 | $\frac{1}{\sqrt{5}}$ | 0 | $\frac{1}{6}\sqrt{\frac{7}{11}}(4+\sqrt{3})$ | $\frac{-2+\sqrt{3}}{3\sqrt{7}}$ | $-\frac{1}{3}\sqrt{\frac{5}{14}}(-2+\sqrt{3})$ | $\frac{4+\sqrt{3}}{3\sqrt{11}}$ |
| | 2 | 3/2 | 3/2 | -1 | 1 | $\sqrt{\frac{2}{35}}$ | 0 | $2\sqrt{\frac{3}{35}}$ | 0 | 0 | $i\sqrt{\frac{2}{5}}-2\sqrt{\frac{3}{5}}$ | $\frac{3i+2\sqrt{6}}{3\sqrt{5}}$ | $2i\sqrt{\frac{2}{5}}-\frac{2}{\sqrt{15}}$ |
| | | | 1/2 | -1 | 1 | $\frac{2}{5\sqrt{7}}$ | $-\frac{2\sqrt{2}}{5}$ | $\sqrt{\frac{2}{35}}$ | 0 | $-\frac{1}{5\sqrt{3}}$ | $-\frac{2}{\sqrt{5}}$ | $-\frac{2}{5\sqrt{3}}$ | $i\sqrt{\frac{2}{5}}$ |
| | | | -1/2 | -1 | 1 | $\frac{-3+2\sqrt{3}}{5\sqrt{7}}$ | $-\frac{2}{15}(3+\sqrt{3})$ | $-\frac{2(-2+\sqrt{3})}{5\sqrt{7}}$ | $\frac{1}{15}\sqrt{2}(3+\sqrt{3})$ | $-\frac{4\sqrt{2}}{15}$ | $-\frac{2\sqrt{\frac{2}{5}}}{3}$ | $-\frac{4}{5\sqrt{3}}$ | 0 |
| | | | -3/2 | -1 | 1 | $\frac{2(-2+\sqrt{3})}{5\sqrt{7}}$ | $-\frac{1}{15}\sqrt{2}(3+\sqrt{3})$ | $\frac{3-2\sqrt{3}}{5\sqrt{7}}$ | $\frac{2}{15}(3+\sqrt{3})$ | 0 | 0 | 0 | 0 |
| | | 5/2 | 5/2 | 1 | 3 | 0 | 0 | 0 | 0 | $-\frac{\sqrt{\frac{5}{14}}}{3}$ | $\frac{3}{\sqrt{70}}$ | $-\frac{2\sqrt{\frac{5}{7}}}{3}$ | $-\frac{1}{2}i\sqrt{\frac{3}{35}}$ |
| | | | 3/2 | 1 | 3 | $\frac{4}{7\sqrt{5}}$ | 0 | $\frac{4\sqrt{\frac{6}{5}}}{7}$ | 0 | $-\frac{\sqrt{\frac{2}{5}}}{7}-\frac{1}{\sqrt{15}}$ | $\frac{-7+6\sqrt{6}}{14\sqrt{5}}$ | $-\frac{6+7\sqrt{6}}{6\sqrt{35}}$ | 0 |
| | | | ½ | 1 | 3 | $\frac{2}{35}(5+\sqrt{3})$ | $-\frac{2\sqrt{\frac{6}{7}}}{5}+\frac{1}{\sqrt{14}}$ | $\frac{1}{7}\sqrt{\frac{2}{5}}(5+\sqrt{3})$ | 0 | $\frac{1}{35}(-5\sqrt{2}-3\sqrt{5})$ | $\frac{1}{35}\sqrt{3}(-4+3\sqrt{10})$ | $-\frac{2}{7}-\frac{3\sqrt{\frac{2}{5}}}{7}$ | $\frac{1}{70}(4\sqrt{3}-3\sqrt{30})$ |
| | | | -1/2 | 1 | 3 | $\frac{3}{35}(3+\sqrt{3})$ | $\frac{2-3\sqrt{3}}{5\sqrt{7}}$ | $\frac{6}{35}(1+\sqrt{3})$ | $\frac{-2+3\sqrt{3}}{5\sqrt{14}}$ | $-\frac{2}{105}(15+\sqrt{10})$ | $-\frac{3}{70}(\sqrt{2}-4\sqrt{5})$ | $\frac{1}{21}(-\sqrt{2}-3\sqrt{5})$ | $\frac{1}{35}(\sqrt{3}-2\sqrt{30})$ |
| | | | -3/2 | 1 | 3 | $\frac{2}{35}\sqrt{2}(-1+3\sqrt{3})$ | $-\frac{1+2\sqrt{3}}{5\sqrt{7}}$ | $-\frac{1}{35}\sqrt{2}(-9+\sqrt{3})$ | $\frac{1}{35}(\sqrt{14}+2\sqrt{42})$ | $\frac{1}{42}\sqrt{5}(-6+\sqrt{6})$ | $\frac{4+3\sqrt{6}}{14\sqrt{5}}$ | $\frac{1}{21}(-6+\sqrt{6})$ | $-\frac{9+2\sqrt{6}}{14\sqrt{5}}$ |
| | | | -5/2 | 1 | 3 | $\frac{2}{7}(-1+\sqrt{3})$ | 0 | $\frac{2}{7}\sqrt{\frac{2}{5}}(-1+\sqrt{3})$ | $\frac{4+\sqrt{3}}{\sqrt{35}}$ | $\frac{\sqrt{5}}{7}$ | $\frac{\sqrt{15}}{14}$ | $\frac{\sqrt{\frac{5}{2}}}{7}$ | $-\frac{\sqrt{15}}{7}$ |
| | | 7/2 | 7/2 | 3 | 5 | 0 | 0 | 0 | 0 | $\frac{1}{55}\sqrt{\frac{2}{39}}(20+\sqrt{5})$ | $\frac{1}{55}\sqrt{\frac{2}{3}}(-5+2\sqrt{5})$ | $\frac{20+\sqrt{5}}{5\sqrt{429}}$ | 0 |
| | | | 5/2 | 3 | 5 | 0 | 0 | 0 | 0 | $\frac{1}{715}(2\sqrt{130}+3\sqrt{1365})$ | $-\frac{4}{165}(\sqrt{30}-\sqrt{35})$ | $\frac{1}{429}(4\sqrt{39}+9\sqrt{182})$ | $\frac{1}{165}(2\sqrt{15}-\sqrt{70})$ |
| | | | 3/2 | 3 | 5 | $\frac{1}{7}$ | 0 | $\frac{\sqrt{6}}{7}$ | 0 | $\frac{2}{55}\sqrt{\frac{2}{13}}(5+4\sqrt{15})$ | $\frac{2}{495}(-10\sqrt{21}+9\sqrt{35})$ | $\frac{3}{55}\sqrt{\frac{2}{13}}(5+4\sqrt{15})$ | $\frac{2}{495}(10\sqrt{6}-9\sqrt{10})$ |
| | | | 1/2 | 3 | 5 | $\frac{1}{35}(\sqrt{2}+4\sqrt{6})$ | $\frac{2(-1+\sqrt{3})}{5\sqrt{7}}$ | $\frac{1}{35}(\sqrt{5}+4\sqrt{15})$ | 0 | $\frac{1}{429}(4\sqrt{195}+5\sqrt{546})$ | $\frac{2}{33}(2\sqrt{5}-\sqrt{14})$ | $\frac{4}{429}(2\sqrt{78}+\sqrt{1365})$ | $\frac{2}{33}(\sqrt{7}-\sqrt{10})$ |
| | | | -1/2 | 3 | 5 | $\frac{1}{35}\sqrt{2}(9+2\sqrt{3})$ | $\frac{2}{5}\sqrt{\frac{2}{7}}(-1+\sqrt{3})$ | $\frac{2}{35}\sqrt{2}(2+3\sqrt{3})$ | $-\frac{2(-1+\sqrt{3})}{5\sqrt{7}}$ | $\frac{12+\sqrt{70}}{11\sqrt{13}}$ | $\frac{1}{33}(10-\sqrt{70})$ | $\frac{1}{429}(7\sqrt{195}+6\sqrt{546})$ | $\frac{2}{33}(-2\sqrt{5}+\sqrt{14})$ |
| | | | -3/2 | 3 | 5 | $\frac{2}{7}\sqrt{\frac{2}{5}}(2+\sqrt{3})$ | $-\frac{2(-3+\sqrt{3})}{3\sqrt{35}}$ | $\frac{1}{7}\sqrt{\frac{2}{5}}(3+2\sqrt{3})$ | $\frac{2}{3}\sqrt{\frac{2}{35}}(-3+\sqrt{3})$ | $\frac{7}{715}(5\sqrt{39}+4\sqrt{65})$ | $\frac{2}{495}(-10\sqrt{42}+9\sqrt{70})$ | $\frac{1}{55}\sqrt{\frac{14}{13}}(15+4\sqrt{15})$ | $\frac{1}{99}\sqrt{14}(-9+2\sqrt{15})$ |
| | | | -5/2 | 3 | 5 | $\frac{1}{7}(4+\sqrt{3})$ | 0 | $\frac{1}{7}\sqrt{\frac{2}{5}}(4+\sqrt{3})$ | $-\frac{2(-1+\sqrt{3})}{\sqrt{35}}$ | $\frac{4(21\sqrt{130}+4\sqrt{1365})}{2145}$ | $\frac{2}{165}(7\sqrt{10}-2\sqrt{105})$ | $\frac{2(21+2\sqrt{42})}{33\sqrt{13}}$ | $\frac{4(-7+\sqrt{42})}{33\sqrt{5}}$ |
| | | | -7/2 | 3 | 5 | $\sqrt{\frac{6}{7}}$ | 0 | $\frac{1}{\sqrt{7}}$ | 0 | $\frac{6(20+3\sqrt{5})}{55\sqrt{13}}$ | $\frac{4(-5+2\sqrt{5})}{55\sqrt{3}}$ | $-\frac{2}{55}\sqrt{\frac{14}{3}}(-5+2\sqrt{5})$ | $\frac{4(20+3\sqrt{5})}{55\sqrt{13}}$ |
| | 3 | 3/2 | 3/2 | -3 | 0 | $-\frac{2}{21}\sqrt{2}(-1+\sqrt{3})$ | $-\frac{1}{7}\sqrt{2}(1+2\sqrt{3})$ | $-\frac{2(-1+\sqrt{3})}{3\sqrt{7}}$ | 0 | $-\sqrt{\frac{2}{3}}$ | $i\sqrt{6}$ | $-2i\sqrt{\frac{2}{3}}$ | $-3$ |
| | | | ½ | -3 | 0 | $\frac{2}{21}(-3+\sqrt{3})$ | $-\frac{2}{35}\sqrt{2}(5+2\sqrt{3})$ | $\frac{2}{21}\sqrt{2}(-3+\sqrt{3})$ | $\frac{1}{35}\sqrt{2}(5+2\sqrt{3})$ | 0 | 1 | $\sqrt{2}$ | $-i\sqrt{2}$ |
| | | | -1/2 | -3 | 0 | $\frac{1}{35}(-9-4\sqrt{3})$ | $\frac{4(-3+\sqrt{3})}{21\sqrt{5}}$ | $\frac{2}{21}(-3+\sqrt{3})$ | $\frac{1}{35}\sqrt{2}(4+3\sqrt{3})$ | 0 | 0 | 0 | 0 |
| | | | -3/2 | -3 | 0 | $-\frac{2}{21}(-1+\sqrt{3})$ | $-\frac{1}{35}\sqrt{2}(4+\sqrt{3})$ | $-\frac{4(-1+\sqrt{3})}{21\sqrt{5}}$ | $\frac{1}{35}(3+4\sqrt{3})$ | 0 | 0 | 0 | 0 |
| | | 5/2 | 5/2 | -1 | 2 | $\frac{\sqrt{\frac{5}{21}}}{2}$ | 0 | $\sqrt{\frac{10}{21}}$ | 0 | 0 | $\frac{2i-3\sqrt{2}}{\sqrt{21}}$ | $\sqrt{\frac{3}{35}}(i+2\sqrt{2})$ | $-\frac{2(-3i+\sqrt{2})}{\sqrt{105}}$ |
| | | | 3/2 | -1 | 2 | $\frac{3+2\sqrt{3}}{21\sqrt{2}}$ | $\frac{1}{14}(\sqrt{2}-3\sqrt{6})$ | $\frac{1}{2\sqrt{7}}+\frac{1}{\sqrt{21}}$ | 0 | $\frac{1}{7\sqrt{5}}$ | $-2\sqrt{\frac{6}{35}}$ | $\frac{\sqrt{\frac{6}{5}}}{7}$ | $i\sqrt{\frac{6}{35}}$ |
| | | | ½ | -1 | 2 | $-\frac{-6+\sqrt{3}}{21\sqrt{2}}$ | $-\frac{2}{35}(1+5\sqrt{3})$ | $\frac{1}{21}(6-\sqrt{3})$ | $\frac{1}{35}(1+5\sqrt{3})$ | $\frac{1}{35}(3\sqrt{3}-2\sqrt{6})$ | $-\frac{1}{5}\sqrt{\frac{3}{7}}(1+3\sqrt{2})$ | $\frac{1}{14}\sqrt{\frac{3}{5}}(-4+3\sqrt{2})$ | 0 |
| | | | -1/2 | -1 | 2 | $\frac{1}{21}\sqrt{\frac{2}{5}}(9-4\sqrt{3})$ | $-\frac{3}{35}\sqrt{2}(3+\sqrt{3})$ | $\frac{1}{42}(9\sqrt{2}-4\sqrt{6})$ | $\frac{6}{35}(1+\sqrt{3})$ | $\frac{3}{70}(4-5\sqrt{2})$ | $-\frac{6+5\sqrt{2}}{5\sqrt{21}}$ | $\frac{1}{35}\sqrt{3}(4-5\sqrt{2})$ | $\frac{\sqrt{\frac{6}{7}}}{5}+\frac{1}{\sqrt{21}}$ |
| | | | -3/2 | -1 | 2 | $\frac{2}{7}-\frac{1}{2\sqrt{3}}$ | $-\frac{1}{35}\sqrt{2}(7+3\sqrt{3})$ | $\frac{4}{7\sqrt{5}}-\frac{1}{\sqrt{15}}$ | $\frac{1}{35}(9+7\sqrt{3})$ | $-\frac{8\sqrt{2}}{35}$ | $-\frac{4}{5\sqrt{7}}$ | $-\frac{4\sqrt{6}}{35}$ | $\frac{4\sqrt{\frac{2}{7}}}{5}$ |
| | | | -5/2 | -1 | 2 | $\frac{1}{21}\sqrt{\frac{5}{2}}(3-2\sqrt{3})$ | $-\frac{5+\sqrt{3}}{7\sqrt{10}}$ | $\frac{1}{42}\sqrt{5}(3-2\sqrt{3})$ | $\frac{1}{7}\sqrt{\frac{2}{5}}(5+\sqrt{3})$ | 0 | 0 | 0 | 0 |
| | | 7/2 | 7/2 | 1 | 4 | 0 | 0 | 0 | 0 | $-\frac{\sqrt{\frac{14}{55}}}{3}$ | $2\sqrt{\frac{2}{105}}$ | $-\frac{2\sqrt{\frac{7}{11}}}{3}$ | $-\frac{i}{\sqrt{105}}$ |
| | | | 5/2 | 1 | 4 | $\frac{2\sqrt{\frac{2}{21}}}{3}$ | 0 | $\frac{8}{3\sqrt{21}}$ | 0 | $-\frac{2(22+3\sqrt{3})}{9\sqrt{385}}$ | $\frac{-11+12\sqrt{3}}{9\sqrt{35}}$ | $-\frac{1}{3}\sqrt{\frac{2}{385}}(22+3\sqrt{3})$ | 0 |
| | | | 3/2 | 1 | 4 | $\frac{2}{63}(9+\sqrt{3})$ | $\frac{1}{7}(-2+\sqrt{3})$ | $\frac{1}{9}\sqrt{\frac{2}{7}}(9+\sqrt{3})$ | 0 | $-\frac{4\sqrt{\frac{2}{55}}}{3}-\frac{2}{\sqrt{231}}$ | $-\frac{8\sqrt{\frac{2}{5}}}{21}+\frac{2}{\sqrt{21}}$ | $-\frac{4\sqrt{\frac{2}{77}}}{3}-\frac{16}{3\sqrt{165}}$ | $\frac{1}{315}(-15\sqrt{14}+8\sqrt{15})$ |
| | | | ½ | 1 | 4 | $\frac{1}{21}(\sqrt{10}+\sqrt{30})$ | $\frac{2(6-5\sqrt{3})}{21\sqrt{5}}$ | $\frac{2}{21}(\sqrt{5}+\sqrt{15})$ | $\frac{-6+5\sqrt{3}}{21\sqrt{5}}$ | $-\frac{4}{693}(3\sqrt{154}+\sqrt{1155})$ | $\frac{1}{63}(12\sqrt{5}-5\sqrt{6})$ | $-\frac{2}{99}(3\sqrt{22}+\sqrt{165})$ | $\frac{2}{63}(-6\sqrt{2}+\sqrt{15})$ |
| | | | -1/2 | 1 | 4 | $\frac{4}{63}\sqrt{2}(3+\sqrt{3})$ | $\frac{1}{7}\sqrt{\frac{2}{5}}(-4+\sqrt{3})$ | $\frac{2}{63}\sqrt{10}(3+\sqrt{3})$ | $\frac{2(-3+4\sqrt{3})}{21\sqrt{5}}$ | $-2\sqrt{\frac{5}{231}}-\frac{5\sqrt{\frac{2}{77}}}{9}$ | $\frac{2}{63}(-2\sqrt{2}+3\sqrt{15})$ | $-\frac{2}{693}(9\sqrt{154}+\sqrt{1155})$ | $\frac{1}{63}(-9\sqrt{5}+2\sqrt{6})$ |
| | | | -3/2 | 1 | 4 | $\frac{10\sqrt{\frac{2}{3}}}{21}$ | $-\frac{2}{7}$ | $\frac{4\sqrt{\frac{10}{3}}}{21}$ | $\frac{\sqrt{6}}{7}$ | $\frac{2\sqrt{\frac{2}{385}}}{3}-\frac{2}{\sqrt{33}}$ | $\frac{1}{21}+2\sqrt{\frac{2}{105}}$ | $\frac{1}{693}(-21\sqrt{110}+2\sqrt{231})$ | $-\frac{2}{315}(5\sqrt{3}+6\sqrt{70})$ |
| | | | -5/2 | 1 | 4 | $-\frac{2}{21}(-5+\sqrt{3})$ | $\frac{1}{21}(-3-2\sqrt{3})$ | $-\frac{1}{21}\sqrt{2}(-5+\sqrt{3})$ | $\frac{2}{21}(3+2\sqrt{3})$ | $\frac{2}{9}\sqrt{\frac{14}{55}}(-3+\sqrt{3})$ | $\frac{2}{9}\sqrt{\frac{2}{35}}(3+2\sqrt{3})$ | $\frac{4}{9}\sqrt{\frac{2}{55}}(-3+\sqrt{3})$ | $-\frac{2(3+2\sqrt{3})}{9\sqrt{7}}$ |
| | | | -7/2 | 1 | 4 | $-\frac{1}{9}\sqrt{2}(-3+\sqrt{3})$ | 0 | $-\frac{2(-3+\sqrt{3})}{9\sqrt{7}}$ | $\frac{2}{\sqrt{7}}+\frac{1}{\sqrt{21}}$ | $\frac{8\sqrt{\frac{7}{55}}}{9}$ | $\frac{2\sqrt{\frac{7}{5}}}{9}$ | $\frac{2\sqrt{\frac{14}{165}}}{3}$ | $-\frac{2\sqrt{\frac{14}{15}}}{3}$ |
| | | 9/2 | 9/2 | 3 | 6 | 0 | 0 | 0 | 0 | $\frac{30\sqrt{22}+\sqrt{165}}{2145}$ | $\frac{1}{143}(\sqrt{55}-\sqrt{66})$ | $\frac{2}{\sqrt{143}}+\frac{1}{\sqrt{4290}}$ | 0 |
| | | | 7/2 | 3 | 6 | 0 | 0 | 0 | 0 | $\frac{33\sqrt{30}+4\sqrt{55}}{1430}$ | $\frac{1}{143}(3\sqrt{30}-2\sqrt{55})$ | $\frac{1}{715}(33\sqrt{30}+4\sqrt{55})$ | $\frac{1}{143}(-3\sqrt{3}+\sqrt{22})$ |
| | | | 5/2 | 3 | 6 | $\frac{1}{6\sqrt{3}}$ | 0 | $\frac{\sqrt{\frac{2}{3}}}{3}$ | 0 | $\frac{10}{13\sqrt{11}}$ | $\frac{3\sqrt{2}}{143}$ | $\frac{5}{13}$ | $-\frac{2}{143}$ |
| | | | 3/2 | 3 | 6 | $\frac{18+\sqrt{3}}{63\sqrt{2}}$ | $\frac{-1+\sqrt{3}}{7\sqrt{2}}$ | $\frac{1}{\sqrt{7}}+\frac{1}{6\sqrt{21}}$ | 0 | $\frac{1}{429}(2\sqrt{330}+3\sqrt{1155})$ | $\frac{4}{143}\sqrt{5}(-3+\sqrt{14})$ | $\frac{1}{13}\sqrt{\frac{5}{33}}(4+3\sqrt{14})$ | $\frac{1}{143}(3\sqrt{30}-2\sqrt{105})$ |
| | | | ½ | 3 | 6 | $\frac{2+5\sqrt{3}}{21\sqrt{2}}$ | $-\frac{2}{21}(-3+\sqrt{3})$ | $\frac{1}{21}(2+5\sqrt{3})$ | $\frac{1}{21}(-3+\sqrt{3})$ | $\frac{12+\sqrt{35}}{13\sqrt{11}}$ | $\frac{1}{143}(-14\sqrt{5}+15\sqrt{7})$ | $\frac{1}{13}\sqrt{\frac{3}{22}}(12+\sqrt{35})$ | $\frac{2}{143}(-15+2\sqrt{35})$ |
| | | | -1/2 | 3 | 6 | $\frac{1}{63}\sqrt{10}(3+4\sqrt{3})$ | $\frac{1}{7}\sqrt{2}(-1+\sqrt{3})$ | $\frac{5(3+4\sqrt{3})}{63\sqrt{2}}$ | $\frac{2}{21}(-3+\sqrt{3})$ | $\frac{7(4\sqrt{110}+5\sqrt{154})}{1430}$ | $\frac{1}{143}\sqrt{6}(-14+3\sqrt{35})$ | $\frac{2(35+4\sqrt{35})}{65\sqrt{11}}$ | $\frac{1}{143}(14\sqrt{5}-15\sqrt{7})$ |
| | | | -3/2 | 3 | 6 | $\frac{5}{126}(9+4\sqrt{3})$ | $\frac{1}{7}\sqrt{2}(-1+\sqrt{3})$ | $\frac{1}{63}\sqrt{5}(9+4\sqrt{3})$ | $\frac{1}{7}(-3+\sqrt{3})$ | $\frac{4(7+2\sqrt{14})}{13\sqrt{55}}$ | $\frac{2}{143}(7\sqrt{10}-3\sqrt{35})$ | $\frac{7(4+\sqrt{14})}{13\sqrt{55}}$ | $\frac{2}{143}\sqrt{3}(-14+3\sqrt{14})$ |
| | | | -5/2 | 3 | 6 | $\frac{5+2\sqrt{3}}{3\sqrt{14}}$ | $\frac{1}{\sqrt{14}}-\frac{1}{\sqrt{42}}$ | $\frac{5}{6\sqrt{7}}+\frac{1}{\sqrt{21}}$ | $\frac{1}{3}\sqrt{\frac{2}{7}}(-3+\sqrt{3})$ | $\frac{21\sqrt{\frac{2}{11}}}{13}$ | $\frac{4\sqrt{3}}{143}$ | $\frac{14\sqrt{\frac{3}{11}}}{13}$ | $-\frac{2\sqrt{21}}{143}$ |
| | | | -7/2 | 3 | 6 | $\frac{1}{9}(6+\sqrt{3})$ | 0 | $\frac{1}{9}\sqrt{\frac{2}{7}}(6+\sqrt{3})$ | $\frac{1}{3}\sqrt{\frac{2}{7}}(-3+\sqrt{3})$ | $\frac{1}{143}\sqrt{5}(33+2\sqrt{66})$ | $-\frac{3}{143}(-3\sqrt{15}+\sqrt{110})$ | $\frac{1}{286}(33\sqrt{10}+4\sqrt{165})$ | $\frac{2}{143}(-9\sqrt{10}+2\sqrt{165})$ |
| | | | -9/2 | 3 | 6 | $\frac{2\sqrt{2}}{3}$ | 0 | $\frac{1}{3}$ | 0 | $\frac{1}{78}(3\sqrt{22}+4\sqrt{165})$ | $\frac{1}{143}(5\sqrt{22}-2\sqrt{165})$ | $\frac{1}{39}(3\sqrt{2}+4\sqrt{15})$ | $\frac{3(-5+\sqrt{30})}{13\sqrt{11}}$ |